\documentclass[superscriptaddress,amsmath,reprint]{revtex4-1}
\usepackage[english]{babel}
\usepackage{graphicx, amsfonts, amssymb}
\usepackage{hyperref}
\usepackage{colortbl}
\usepackage{siunitx}
\usepackage{mathtools}

\newcommand{\eqr}[1]{Eq.~\eqref{#1}}
\newcommand{\cit}[1]{Ref.~\cite{#1}}
\newcommand{\fig}[1]{Fig.~\ref{#1}}

\newcommand{\nn}{\nonumber\\}
\newcommand{\ds}[1]{\displaystyle{#1}}

\newcommand{\kBT}[0]{k_\mathrm{B}T}
\newcommand{\kbt}{k_{\rm B}T}

\newcommand{\rhos}{\rho_{\rm s}}

\begin{document}

\author{Cheng Lian}
\email{c.lian@uu.nl}
\affiliation{Institute for Theoretical Physics, Center for Extreme Matter and Emergent Phenomena, Utrecht University, Princetonplein 5, 3584 CC Utrecht, The Netherlands}
\author{Mathijs Janssen}
\affiliation{Institute for Theoretical Physics, Center for Extreme Matter and Emergent Phenomena, Utrecht University, Princetonplein 5, 3584 CC Utrecht, The Netherlands}
\affiliation{Max-Planck-Institut f\"{u}r Intelligente Systeme, Heisenbergstra{\ss}e 3, 70569 Stuttgart, Germany}
\affiliation{Institut f\"{u}r Theoretische Physik IV, Universit\"{a}t Stuttgart, Pfaffenwaldring 57, 70569 Stuttgart, Germany}
\author{Honglai Liu}
\affiliation{State Key Laboratory of Chemical Engineering, School of Chemistry and Molecular Engineering, East China University of Science and Technology, Shanghai 200237, China}
\author{Ren\'{e} van Roij}
\affiliation{Institute for Theoretical Physics, Center for Extreme Matter and Emergent Phenomena, Utrecht University, Princetonplein 5, 3584 CC Utrecht, The Netherlands}
\date{\today}

\begin{abstract}
The development of novel electrolytes and electrodes for supercapacitors is hindered by a gap of several orders of magnitude between experimentally measured and theoretically predicted charging timescales.
Here, we propose an electrode model, containing many parallel stacked electrodes, that explains the slow charging dynamics of supercapacitors.
At low applied potentials, the charging behavior of this model is described well by an equivalent circuit model.
Conversely, at high potentials, charging dynamics slow down and evolve on two relaxation time scales: a generalized $RC$ time and a diffusion time, which, interestingly, become similar for porous electrodes. 
The charging behavior of the stack-electrode model presented here helps to understand the charging dynamics of porous electrodes and qualitatively agrees with experimental time scales measured with porous electrodes.
\end{abstract}

\title{A Blessing and a Curse: How a Supercapacitor's Large Capacitance Causes its Slow Charging}
\maketitle

\begin{figure}[t!]
\includegraphics[width=8.0cm]{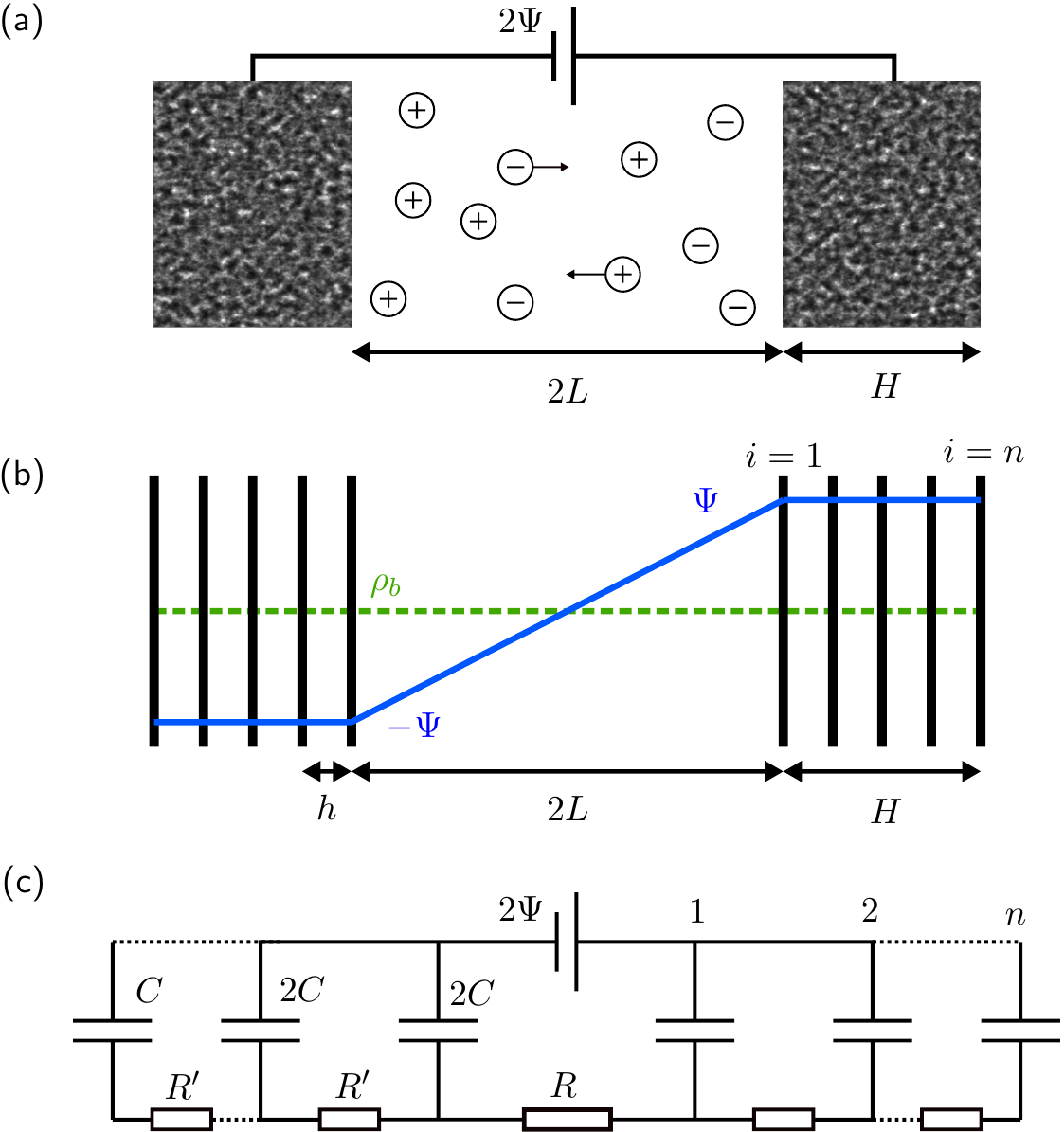}
\caption{(a) Sketch of a supercapacitor containing a 1:1 electrolyte, two porous electrodes, and a battery providing an electrostatic potential difference $2\Psi$. (b) In our stack-electrode model, the cathode and anode each contain $n$ planar electrodes at intervals of $h$. 
Initial anionic and cationic densities are $\rho_b$ throughout the cell. At time $t=0$, $-\Psi$ and $+\Psi$ are applied to all electrodes on the left and right-hand sides of the system, respectively. \mbox{(c) Equivalent} circuit model for the stack-electrode model. }
\label{fig1}
\end{figure}

In the electric energy storage domain, supercapacitors [\fig{fig1}(a)] have proven their value in applications requiring higher power output than delivered by batteries and more energy than stored in dielectric capacitors \cite{chmiola2006sci, forse2017direct, prehal2018salt, limmer2013charge}. 
Many types of carbon-based materials have been used for the capacitor's electrodes \cite{simon2012capacitive, mouterde2019molecular,cheng2016ion}.
However, the relation between the porous structures and the charging dynamics of macroscopic supercapacitors is poorly understood.
On the one hand, transmission line (TL) models \cite{DELEVIE1963751, PhysRevLett.113.097701, tivony2018charging, HELSETH2019} can successfully fit experimental data, but the fit parameters therein do not have a direct interpretation in terms of microscopic properties of supercapacitors.
On the other hand, molecular dynamics simulation \cite{feng2011supercapacitor, kondrat2014NM, pean2014dynamics, breitsprecher2018charge, Guang2019PRX, C8CP07200K, bi2019molecular}, lattice Boltzmann simulations \cite{Benjamin2019jcp, chatterji2007lb}, and classical dynamic density functional theory \cite{Lian2016JCP, babel2018impedance,jiang2014kinetic} can elucidate the charging mechanisms of a single or a few nanopores or a nanoscale anode-cathode model, but predicted relaxation time scales are of the order of $\si{\nano\second}$, roughly 12 orders of magnitude smaller than experimentally measured $10^{3}~\si{\second}$ timescales of supercapacitors \cite{brogioli2011prototype, lian2016enhancing, janssen2017coulometry, Ambrozevich2018}. These long charging times, however, can be decently approximated by  multiplying a nanocapacitor's $RC$ time by the squared ratio of nanocapacitor-to-supercapacitor
thicknesses \cite{pean2014dynamics, bi2019molecular}. Even though such approaches to bridge scales are valuable, they ignore the actual multiscale character of the system, e.g. the transport of ions through quasi-neutral macropores. To faithfully describe the charging of supercapacitors, one should account for both the ionic currents from the ion reservoir that separates anode and cathode through a macropore network into the nanopores (micrometers) and for the electric double layer (EDL) buildup therein (nanometers).
Clearly, such a multiscale analysis cannot be performed with the above mentioned simulation techniques alone, as computational power limits simulations to nanoscale systems.
In this article, we present a minimal model to explain the long experimental relaxation timescales of supercapacitors instead.

The canonical model describing ionic charge relaxation to an applied electric field employs a dilute 1:1 electrolyte and two parallel and planar blocking electrodes separated by a distance $2L$ \cite{bazant2004diffuse}.
Suddenly applying a potential difference $2\Psi$ with a battery, the two electrodes will acquire opposite surface charge densities $\pm e\sigma(t)$, eventually screened in the electrolyte by two EDLs, whose equilibrium width is characterized by the Debye length $\kappa^{-1}=\sqrt{\varepsilon\kBT/2e^2\rho_{b}}$, with $2\rho_{b}$ the bulk ion number density, $\varepsilon$ the electrolyte permittivity, $e$ the elementary charge, and $\kBT$ the thermal energy.
At late times and for $\kappa L\gg1$, $\sigma(t)= 2\Phi\rho_b\kappa^{-1} \left[1-\exp{\left(-t/\tau_{RC}\right)}\right]$, with $2\Phi= 2e\Psi/\kbt$ the dimensionless applied potential, $\tau_{RC}=\kappa^{-1}L/D$ the $RC$ time, and $D$ the ionic diffusion coefficient \cite{bazant2004diffuse, janssen2018transient, Palaia2019}.
Inserting typical experimental parameters $\kappa^{-1}\approx1~\si{\nano\meter}$, $L\approx250~\si{\micro\meter}$, and \mbox{$D\approx1\times10^{-9}~\si{\meter\squared\per\second}$} yields $\tau_{RC}\approx 10^{-4}~\si{\second}$: larger than the timescales predicted by molecular simulations, but still 5 orders of magnitude smaller than the experimental charging time of supercapacitors. 
This discrepancy comes as no surprise as the above $\sigma(t)$ applies to planar electrodes: this model does not account for the huge surface area and for the ion transport through the porous structure of the supercapacitor electrodes.
Simple extensions of the flat electrode setup were discussed, such as spherical and cylindrical electrodes \cite{Benjamin2019jcp, janssen2019II} and a single cylindrical pore in contact with a reservoir \cite{PhysRevLett.113.097701}.
Several theoretical works focused on the charging dynamics of porous electrodes \cite{pilon2015recent, zhao2013optimization, cheng2019acsnano, PhysRevE.76.011501, PhysRevE.81.031502, Benjamin2014pre, Ali2019}.  Still, the gap between experimental and theoretical supercapacitor charging timescales has not been bridged yet.

To explain why the charging time of macroscopic porous electrodes [\fig{fig1}(a)] is much larger than that of flat electrodes, in this article, we will characterize the charging dynamics of the model shown in \fig{fig1}(b) and also compare it to the charging dynamics of the circuit shown in \fig{fig1}(c).
In our model, the nanoporous cathode and anode of a supercapacitor are both modeled by a stack of $n$ parallel electrodes with an equal spacing $h$ mimicking the pore size, such that the thickness of the cathode and anode equals $H=(n-1)h$. The surface area $A$ of all individual electrodes is assumed to be sufficiently large that we can ignore edge effects and study all microscopic observables as a function of a single coordinate $x$ perpendicular to the electrode surfaces.
We adopt a coordinate system whose origin lies in the middle ($x=0$) of the system and where the $i$th cathode and anode, with $i=\{1,\ldots,n\}$, are located at $X_i=\pm [L+(i-1)h]$.
All parallel stacked electrodes are fully permeable to the electrolyte in order to mimic the porosity of supercapacitor electrodes, except the two outer ones $(i=n)$ which are impermeable to have a closed system [cf. \eqr{eq:eln}].
Thus, the ionic number densities $\rho_{\pm}(x,t)$ and ionic fluxes $ j_{\pm}(x,t)$ are continuous at each $X_{i}$.
Initially, the ionic number densities are homogenous
\begin{align} \label{eq:1}
\rho_{\pm}(x,t=0)=\rho_b, ~~~ |x|\leq L+H.
\end{align} 	
At time $t=0$, a dimensionless potential difference $2\Phi$ is applied to the macroscopic cathode and anode.
This yields the following boundary conditions for $t>0$,
\begin{subequations}\label{eq:2}
\begin{align}
\phi(\pm X_i,t)&=\pm\Phi; \label{eq:BCpot}\\
j_{\pm}(\pm X_n,t)&=0\,, \label{eq:eln}
\end{align} 	 		
\end{subequations}
with $\phi(x,t)$ the electric potential in units of the thermal voltage $\kBT/e$.
To model the ionic dynamics, we use the classical Poisson-Nernst-Planck (PNP) equations \cite{bazant2004diffuse}
\begin{subequations}\label{eq:3}
\begin{align}
\partial^2_{x} \phi(x,t)&=-\kappa^2\left[\frac{\rho_{+}(x,t)-\rho_{-}(x,t)}{2\rho_b}\right];\\
\partial_{t} \rho_{\pm}(x,t)&=-\partial_{x} j_{\pm}(x,t); \\
j_{\pm}(x,t)&= -D \left[\partial_{x} \rho_{\pm}(x,t) \pm \rho_{\pm}(x,t)\partial_{x} \phi(x,t) \right]\,.
\end{align}
\end{subequations}
In Eqs.~\eqref{eq:1}--\eqref{eq:3} appear the applied potential $\Phi$ and four length scales: $h$, $H$, $L$, and $\kappa^{-1}$. With these parameters, we can construct many different combinations that yield 1+3 independent dimensionless parameters, for instance: $\Phi$, $\kappa L$, $\kappa H$, and $\kappa h$ or, equivalently, $\Phi$, $\kappa L$, $H/L$, and $n$.
Here, we focus on the latter choice and mostly restrict to $H/L=1$, which is reasonable for supercapacitors.

\begin{figure}[b]
\includegraphics[width=8.8cm]{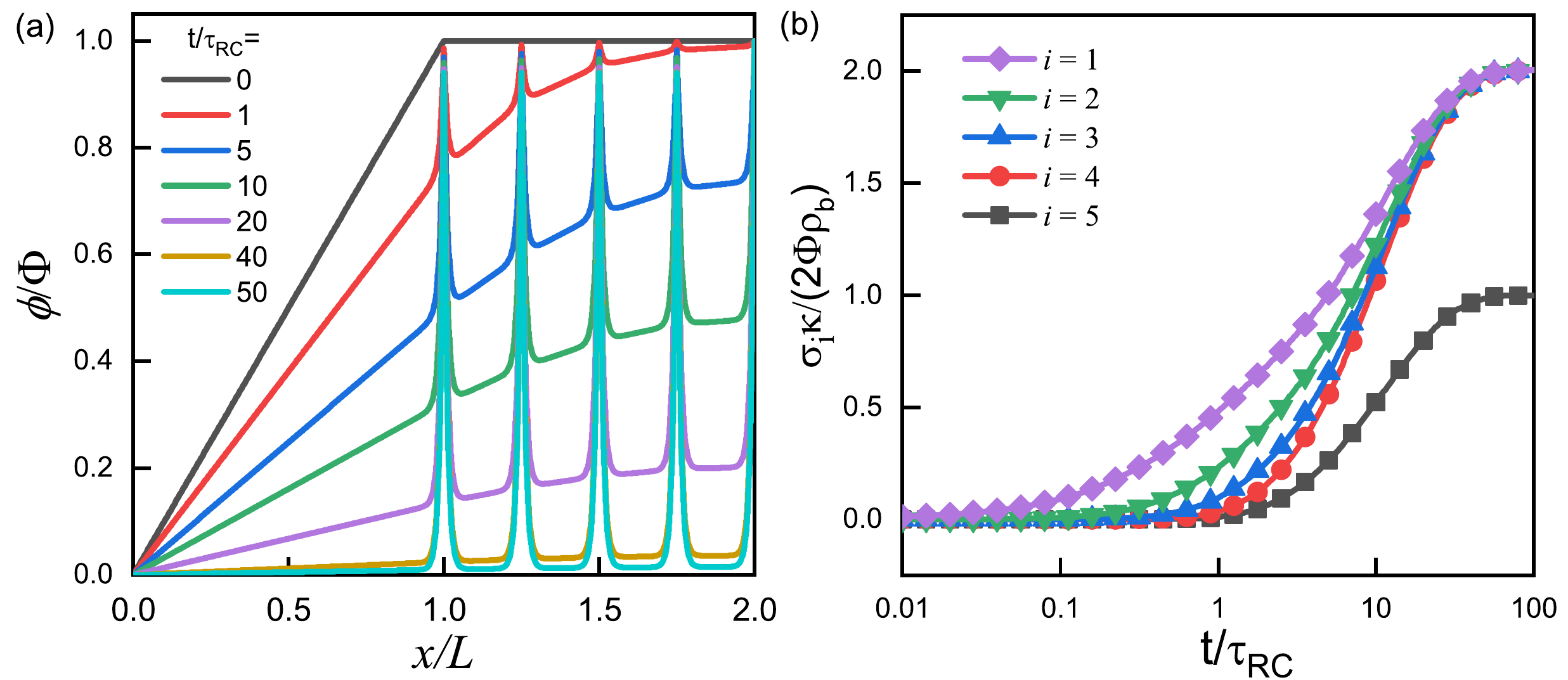}
\caption{Time dependence of the scaled (a) potential $\phi(x,t)$ and (b) the surface charge densities $\sigma_{i}(t)$ for $i=\{1,..,5\}$ of the electrode for $\Phi=0.001$, $\kappa L= 100$, $H/L=1$, and $n=5$. The symbols and lines in (b) correspond to numerical and equivalent-circuit model calculations, respectively.}
\label{fig2}
\end{figure}

In \fig{fig2}(a), we present numerical results for $\phi(x,t)$ for a low potential $\Phi = 0.001$ and $\kappa L = 100$, $H/L=1$, and $n = 5$.
These parameters correspond to $\kappa h = 25$, which means that the EDLs are thin compared to the electrode separations.
Initially, the potential in the reservoir $(|x| < L)$ displays a typical linear $x$ dependence, which corresponds to a spatially constant electric field.
At later times, the potential retains this linear dependence in the reservoir, albeit with a slope that decreases with time due to the buildup of EDLs.
At short times $t/\tau_{RC}\leq20$, there is a clear asymmetry between the two EDLs that surround individual planar electrodes.
This asymmetry is lost at later times $t \geq 20\tau_{RC}$, when the EDLs fully equilibrate.

For the same parameters, in \fig{fig2}(b) we show the surface charge densities $e\sigma_{i}$ of the individual electrodes (labeled with $i$), which we find with Gauss' law $\sigma_{i}(t)=-2\rho_b \kappa^{-2}[\partial_{x}\phi|_{X_{i}^+}-\partial_{x}\phi|_{X_{i}^-}]$.
At early times $t< \tau_{RC}$, the electrodes charge faster the closer they are situated to the reservoir, $\sigma_{1}>\sigma_{2}>...>\sigma_{n}$.
However, the late-time relaxation timescale is the same for all electrodes: all electrodes reach $99.9\%$ of their equilibrium charge around $t/\tau_{RC}\approx 50$.
As the outer electrodes face the electrolyte only at one side, we have \mbox{$\partial_{x}\phi|_{X_{n}^+}=\partial_{x}\phi|_{-X_{n}^-}=0$}, and $\sigma_{n}(t/\tau_{RC}\to\infty)$ is a factor of two smaller than the other electrodes.
To better understand these phenomena, we studied the behavior of the circuit model shown in \fig{fig1}(c).
Note the great similarity of this model to traditional TL models used for fitting experimental data: the only difference is the capacitance $C$ of the outermost capacitor, rather than $2C$ in the TL model (see also Appendix~\ref{TLappendix}).
However, in contrast to the TL model, where $R$, $C$, and $n$ are fit parameters, the elements of the circuit in \fig{fig1}(c) are all one-to-one related to electrolyte and electrode properties of our microscopic model, $R=2L/(A \varepsilon \kappa^2 D)$ and $C=A \varepsilon \kappa$.
In Appendix~\ref{Appendix_A} we derive a matrix differential equation [cf.~Eq.~\ref{eq:matrixn}] that relates the potential drops over the $n$ capacitors  to the currents through the $n$ resistors.
With the solution to this equation, we find predictions for the time-dependent charge on each capacitor in this circuit, which translates into a prediction for $\sigma_{i}(t)$ in the corresponding microscopic model, shown in \fig{fig2}(b) with symbols.
Clearly, the predictions from the microscopic and circuit model are indistinguishable.
In line with our earlier observation, the equivalent-circuit model predicts that all electrodes relax exponentially at late times with the same time constant $\tau_{n}$ 
\begin{equation}\label{eq:4}
\frac{\tau_{n}}{\tau_{RC}}= \left(2+0.75\frac{H}{L}\right)n-1-0.91\frac{H}{L}\,,
\end{equation}
which correctly reduces to $\tau_{1}=\tau_{RC}$ for $n=1$ (for which $H/L=0$).
The coefficients $2$ and $1$ appearing  in \eqr{eq:4}  are analytical results obtained in the limit $H/L\to0$;
the other numerical factors relate to the smallest eigenvalue of the almost-Toeplitz matrix in the afore-mentioned matrix differential equation \ref{eq:matrixn}.
From \eqr{eq:4} we see that $\tau_n$ is large whenever $n$ is large.
This suggests that the large relaxation time of supercapacitors stems from their large internal surface area, achieved through many small pores.
Interestingly, \eqr{eq:4} recovers the electronic circuit intuition that a supercapacitor should charge slower the larger its electrode surface area. This areal scaling is not present in the relaxation time $\tau_{RC}$ of \cit{bazant2004diffuse} as, in that work, $R$ and $C$ scale oppositely with $A$.
\begin{figure}
\includegraphics[width=8.5cm]{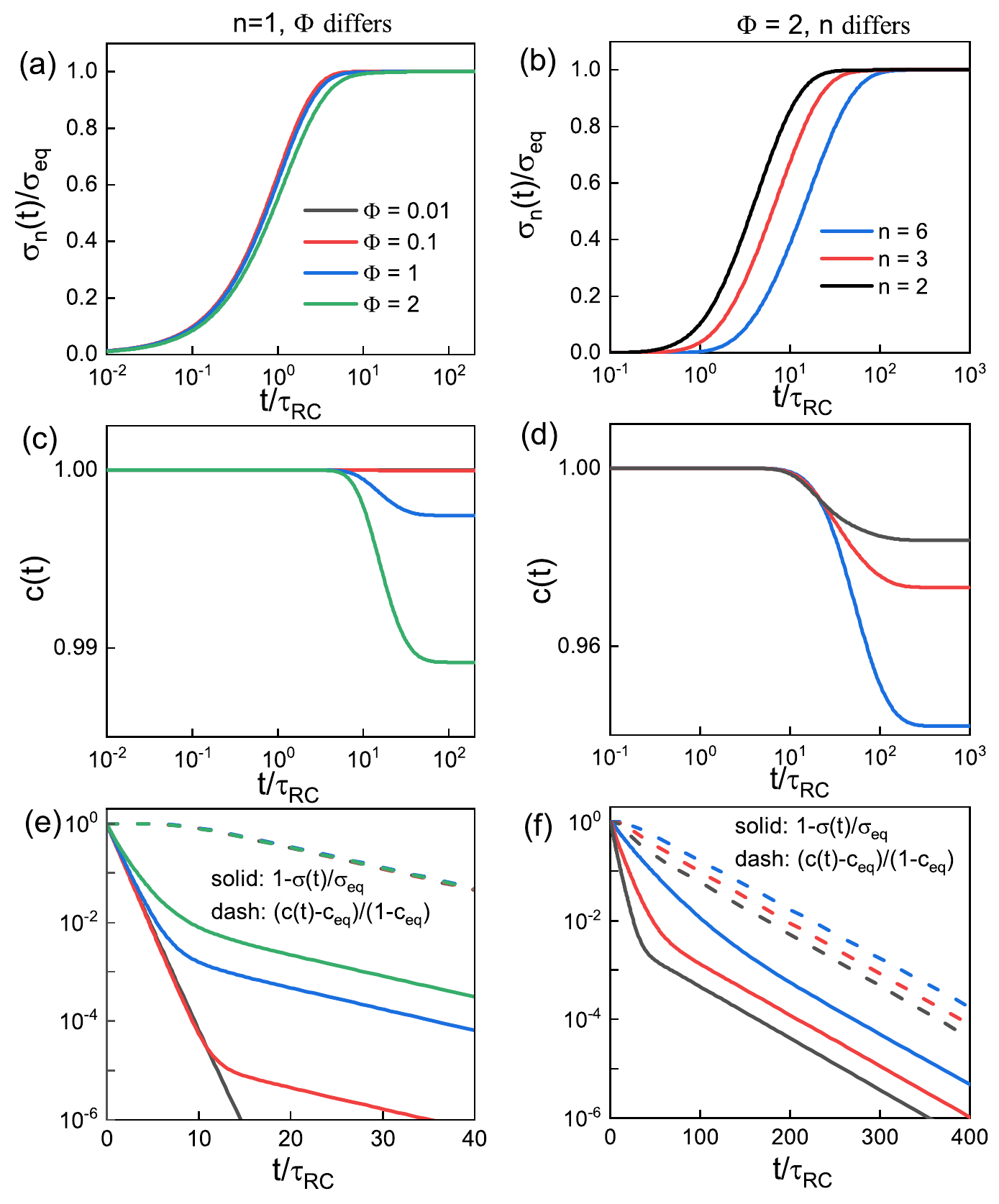}
\caption{ The surface charge density (a), (b) $\sigma(t)/\sigma_{eq}$, the salt concentration at the cell center (c), (d) $c(t)$, and the charge relaxation (e), (f) $1-\sigma(t)/\sigma_{eq}$ (solid lines) and the concentration decay $[c(t)-c_{eq}]/[c(0)-c_{eq}]$ (dashed lines) of the stack-electrode model for $\kappa L=100$, in (a), (c), and (e) for $n=1$, $H/L=0$ at potentials $\Phi=\{0.01, 0.1, 1, 2\}$, and in (b), (d), and (f) for $\Phi=2$ and $H/L=1$, at $n=\{2,3,6 \}$.
}
\label{fig3}
\end{figure}

Because supercapacitors are typically subjected to large potentials in practical applications, we also characterize the dynamics of the stack-electrode model at $\Phi>1$.
In Figs.~\ref{fig3}(a), \ref{fig3}(c), and \ref{fig3}(e) we present data for $\Phi=\{0.01, 0.1, 1, 2\}$, $\kappa L=100$, and $n=1$ (hence, $H/L=0$).
For this two-electrode setup, it is known that, next to $\tau_{RC}$, the diffusion time $L^2/D$ emerges in the ionic relaxation due to slow salt diffusion from the cell center to the electrode surfaces \cite{bazant2004diffuse, Palaia2019, PhysRevE.82.011501}.
Indeed, \fig{fig3}(a) shows that the normalized surface density $\sigma_n(t)/\sigma_{eq}$, with $\sigma_{eq}\equiv\sigma_n(t/\tau_{RC}\to\infty)$ the late-time surface charge density, develops slower at higher $\Phi$.
Next, \fig{fig3}(c) shows the salt concentration at the cell center $c(t)=[\rho_{+}(0,t)+\rho_{-}(0,t)]/(2\rho_b)$ for the same $\Phi$.
We see that $c(t)\approx 1$ for $\Phi \leq 0.1$ and that $c(t)$ decreases at late times ($t/\tau_{RC}>10$) by 0.25\% and 1\% for $\Phi=1$ and $2$:
As our setup is closed, a net ionic adsorption on the electrodes ``desalinates" the cell center \cite{boon2011blue}.
To investigate the emergence of the slow timescale at large applied potentials, in Fig~\ref{fig3}(e), we show the charge relaxation $1-\sigma(t)/\sigma_{eq}$ (solid lines) and the concentration decay $[c(t)-c_{eq}]/[1-c_{eq}]$ (dashed lines), where $c_{eq}\equiv c(t/\tau_{RC} \to \infty)$.
At early times ($t/\tau_{RC}<1$), all data for $1-\sigma(t)/\sigma_{eq}$ collapse onto the $\Phi=0.01$ curve, indicating that the initial ionic relaxation is described well by the equivalent circuit model, even for higher $\Phi$.
Thereafter, a second, slower relaxation emerges in $1-\sigma(t)/\sigma_{eq}$, emerging more dominantly for higher $\Phi$.
At late times, the slopes of $1-\sigma(t)/\sigma_{eq}$ and $[c(t)-c_{eq}]/[1-c_{eq}]$ are the same.
Numerical results  [cf. \fig{fig_S7}(d)]  
for the adsorption timescale $\tau_{\rm ad}$, the inverse of these slopes, show that $\tau_{\rm ad}/\tau_{RC}$ is independent of $\Phi$ (for all $\Phi$ considered) and scales linearly with $\kappa L$.
Using the definition of $\tau_{RC}$, we then recover the $L^2/D$ scaling of $\tau_{\rm ad}$ suggested by \cit{bazant2004diffuse, Palaia2019}. 
In Appendix~\ref{Appendix_C}, we check the robustness of our findings with dynamical density functional theory calculations of a room temperature ionic liquid at the experimentally realistic voltage $\Psi=0.5$ V: \mbox{Figure \ref{fig_S10}} shows that the surface charge again relaxes with two distinct relaxation times.

To investigate the effect of $n>1$ for high potentials, in Figs.~\ref{fig3}(b), \ref{fig3}(d), and \ref{fig3}(f), we plot the same observables as in Figs.~\ref{fig3}(a), \ref{fig3}(c), and \ref{fig3}(e), now for $\Phi=2$, $\kappa L=100$, $H/L=1$, and $n=\{2,3,6\}$.
Similar to our $\Phi=0.001$ findings [condensed in $\tau_n$ of \eqr{eq:4}], we see that the charging dynamics at $\Phi=2$ also slows down with increasing $n$.
The salt concentration at the cell center $c(t)$ [\fig{fig3}(d)] is again unaffected at early times $t/\tau_{RC}<5$, after which it decays to an equilibrium value that decreases with $n$.
Thus, our model recovers the intuition that, for two electrodes of the same volume, the one with more pores (and, hence, a large surface area) desalinates an electrolyte reservoir more.
In \fig{fig3}(f), we see that the surface charge again decays on two distinct timescales.
Plotting the same data with time scaled by $\tau_{n}$ instead of by $\tau_{RC}$ [cf. \fig{fig_S8}(c)], all $1-\sigma(t)/\sigma_{eq}$ collapse for $t\leq \tau_{n}$, which shows that the circuit model decently describes the early-time relaxation at high potentials as well. 
Conversely, at late times, we see in \fig{fig3}(f) that $1-\sigma(t)/\sigma_{eq}$ and $[c(t)-c_{eq}]/[1-c_{eq}]$ decay exponentially with a time constant $\tau_{\rm ad}$ that does not depend on $n$ for the parameter set under consideration.
\begin{figure}
\includegraphics[width=8.6cm]{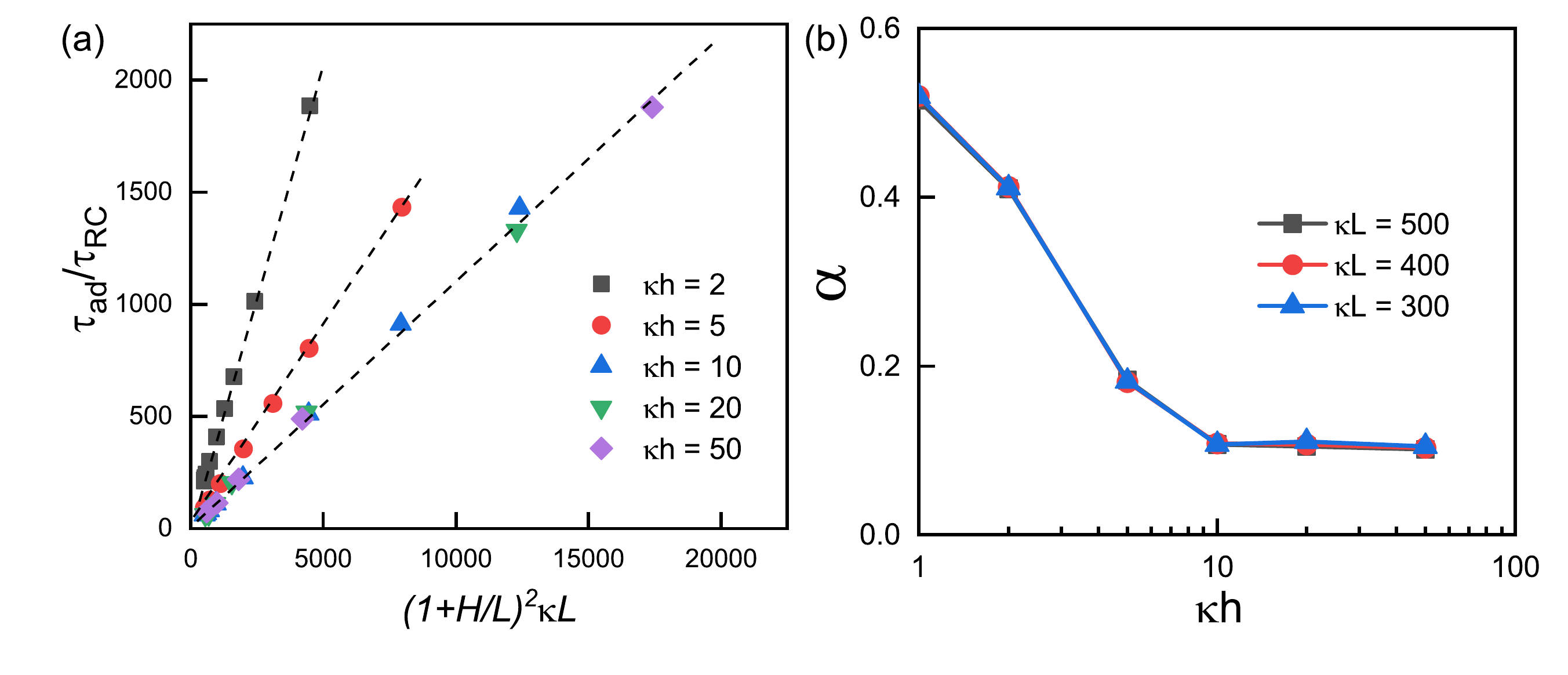}
\caption{(a) The dependence of the (scaled) adsorption timescale $\tau_{\rm ad}$ on the (scaled) system size $(L+H)^2$ for $\kappa L=500$ at potential $\Phi=2$ at several pore sizes $\kappa h=\{2, 5, 10, 20, 50\}$ and $n= \{2, 5, 10, 20, 50, 100, 150, 200, 300\}$, with $n=200$ and $n=300$ for $\kappa h= 20$ and $\kappa h= 50$ off the scale of the plot. (b) The prefactor $\alpha$ of [\eqr{eq:5}] as a function of $\kappa h$ for $\kappa L=500$ with a linear fit through the data in (a); $\kappa L=300$ and $400$ with identical results.
}
\label{fig4}
\end{figure}
Considering a larger set of  $\kappa L$, $\kappa h$, and $n$, we show $\tau_{\rm ad}$ in \fig{fig4}(a). From this figure we conclude that
\begin{equation}\label{eq:5}
\tau_{\rm ad}=\alpha \frac{(H+L)^{2}}{D},
\end{equation}
hence, the adsorption time $\tau_{\rm ad}$  depends on the total system size. We show the $\kappa h$-dependent prefactor $\alpha$ in \fig{fig4}(b) for various $\kappa L$, which reveals that $\alpha$ is $\kappa L$ independent and that $\alpha\approx0.1$ for $\kappa h>10$, while $\alpha$ increases with decreasing $\kappa h\leq10$.

Since $\tau_{n}$ [\eqr{eq:4}] and $\tau_{\rm ad}$ [\eqr{eq:5}] depend on $\kappa L$, $\kappa h$, and $n$ differently, both $\tau_{\rm ad}/\tau_n\ll1$, $\tau_{\rm ad}/\tau_n\approx1$, and $\tau_{\rm ad}/\tau_n\gg1$ are possible.
Focusing here on $n\gg1$, which is relevant to macroscopic electrodes, we find
\begin{equation}\label{eq:an}
\frac{\tau_{n}}{\tau_{\rm ad}}\approx
\begin{dcases}
                        \frac{0.75}{\alpha \kappa h}\frac{H}{L} \quad& L \gg H;\\
                        \frac{0.69}{\alpha \kappa h}    \quad& L = H;\\
                        \frac{2}{\alpha \kappa h}\frac{H}{L}   \quad& L \ll H.
\end{dcases}
\end{equation}
For $H/L=1$ (and $n\gg1$) we find that $\tau_{n}/\tau_{\rm ad}\sim 1$ whenever $\kappa h<10$.

Finally, it is interesting to determine the applicability of our stack-electrode model to experiments:
Here, we consider the setup of \cit{janssen2017coulometry}, where two carbon electrodes of thickness $H=0.5~\si{\milli\meter}$, separation $2L=2.2~\si{\milli\meter}$,  porosity $p=0.65$, mass density $\varrho=5.8\times10^{5}~\si{\gram\per\meter\cubed}$, and Brunauer-Emmett-Teller-area $A_{\rm BET}=1330~\si{\meter\squared\per\gram}$ were used.
Assuming each porous electrode to consist of two flat solid carbon slabs, we get a crude estimate for the pore size with $h=p/(\varrho A_{\rm BET})=0.84~\si{\nano\meter}$.  
The electrodes were immersed in a 1 M NaCl solution at room temperature, hence, $\kappa^{-1}= 0.3~\si{\nano\meter}$ and bulk diffusivity $D=1.6\times10^{-9}~\si{\meter\squared\per\second}$.
[We ignore that $D$ is smaller in nanopores \cite{PhysRevE.76.011501, ALLAHYAROV201773} and that different diffusivities may appear in Eqs.~\eqref{eq:4} and \eqref{eq:5} \cite{janssen2019transient}.]
These parameters correspond, in our model, to $H/L=0.45$, $n=5.9\times10^{5}$, and $\kappa h=2.8$, hence, $\alpha=0.3$.
With Eqs.~\eqref{eq:4} and \eqref{eq:5} we now find $\tau_n=2.9\times10^{2}~\si{\second}$ and $\tau_{\rm ad}=4.8\times10^{2}~\si{\second}$, roughly within 1 order of magnitude from the two timescales ($2\times10^2~\si{\second}$  and $9\times10^3~\si{\second}$) observed in the experimental data of \cit{janssen2017coulometry} (see Appendix~\ref{Appendix_D}).
Given the simplicity of our model and crudeness of our estimates of $\kappa h$, $n$, and $D$, the remaining discrepancies are not surprising.
Yet, the stack-electrode model has bridged the 5-orders-of-magnitude gap between experimental relaxation times and those predicted in the $n=1$ model.

In summary, we studied the charging dynamics of nanoporous electrodes with a simple electrode model.
At small applied potentials, numerical simulations of the PNP equations are reproduced accurately by an equivalent circuit model.
This circuit model is akin to TL models used often to fit experimental supercapacitor data.
Notably, however, the resistances, capacitances, and number of branches in the circuit model are not fit parameters but physically determined by our microscopic model.
This one-to-one relation allows us to interpret the long relaxation time of supercapacitors as being due to the large number $n$ of pores in nanoporous electrodes: The stack-electrode model relaxes with the timescale $\tau_{n} \sim (2+0.75H/L)n\tau_{RC}$.
At higher potentials, the surface charge still relaxes at early times with $\tau_{n}$. 
Higher potentials also lead to slow salt adsorption in the EDLs and concomitant depletion of the reservoir on the timescale $\tau_{\rm ad} \sim (L+H)^2/D$.
As salt and charge transport are coupled, the long timescale $\tau_{\rm ad}$ also governs the late-time surface charge relaxation, all the more so the higher the applied potential.
The two timescales $\tau_n$ and $\tau_{\rm ad}$ differ orders of magnitude for small $n$ but become similar when electrodes have many pores, as is the case for supercapacitors. 
Inserting parameters relating to a recent experimental study \cite{janssen2017coulometry}, our simple model predicts the two observed relaxation times roughly within 1 order of magnitude.
Our model thus successfully bridged the 5-orders-of-magnitude gap between theoretically predicted and experimentally measured timescales, and could serve as a basis for extensions that break the planar symmetry. However, more work is needed to fully understand the charging dynamics of porous electrodes, which should include effects due to finite ion sizes, more realistic modeling of pore morphology, Faradaic reactions, position-dependent diffusion coefficients, etc. 

\begin{acknowledgments}
C.L. and M.J. contributed equally to this work. This work is part of the D-ITP consortium, a program of the Netherlands Organisation for Scientific Research (NWO) that is funded by the Dutch Ministry of Education, Culture and Science (OCW). C.L. and R. v. R. acknowledge the EU-FET project NANOPHLOW (REP-766972-1), M.J. acknowledges support from S. Dietrich, and H.L. acknowledges NSFC (Grants. No. 91834301 and No. 21808055). We kindly thank Pieter Kouyzer for helpful discussions, Ben Ern\'{e} for access to the experimental data of \cit{janssen2017coulometry}, and Sviatoslav Kondrat for comments on our manuscript.
\end{acknowledgments}

\setcounter{figure}{0}
\renewcommand{\thefigure}{{A}\arabic{figure}}
\renewcommand{\theHfigure}{A\arabic{figure}}

\begin{appendix}
\section{Equivalent circuit model}\label{Appendix_A}

We discuss an equivalent circuit model that describes the charging dynamics of the stack-electrode model well (see also \cit{kouyzer2015charging}).
Before turning to the general $n$ case, we first discuss the easier $n=1$ and $n=2$ cases.

\subsection{The case $n=1$}
\begin{figure}[h]
\includegraphics[width=5.0cm]{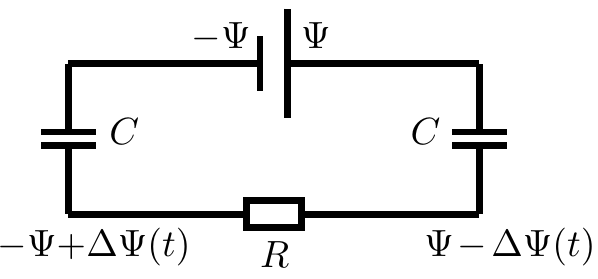}
\caption{Equivalent circuit model for the stack electrode model in the trivial case $n=1$.}\label{fig_S1}
\end{figure}

The trivial $n=1$ case can be described by the circuit model shown in \fig{fig_S1}.
The EDLs on each electrode are modeled with a capacitor of capacitance $C$.
The electrolyte resistance is modeled through a resistor of resistance $R$.
At $t=0$ a voltage source is switched on to provide a potential difference of $2\Psi$.
The capacitors, initially uncharged, will acquire a charge $Q(t)=C\Delta\Psi(t)$
with $\Delta \Psi(t)$ being the time-dependent voltage difference between either sides of the capacitors.
The current that flows through the system is found via Ohm's law,
$IR=[\Psi-\Delta\Psi(t)]-[-\Psi+\Delta\Psi(t)]=2[\Psi-\Delta\Psi(t)]$.
With $I=\dot{Q}(t)=C\Delta \dot{\Psi}(t)$, we find 
\begin{subequations}
\begin{eqnarray}
\Delta \dot{\Psi}(t)&=&[\Psi-\Delta\Psi(t)]\frac{2}{RC} \,,\\
\Delta\Psi(t=0)&=&0\,,
\end{eqnarray}
\end{subequations}
which is solved by 
\begin{eqnarray}\label{seq:nis1}
\Delta\Psi(t)=
\Psi\left[1-\exp{\left(-\frac{2t}{RC}\right)}\right]\,.
\end{eqnarray}
To relate this result to the relaxation of EDLs that we are trying to mimic, note that the EDL capacitance and electrolyte resistivity read $C=A \epsilon \kappa $ and $R=2L /(A \epsilon  \kappa^2 D) $, respectively; hence, $RC =2\tau_{RC} =2\kappa^{-1} L/D $.

\subsection{The case $n=2$}
\begin{figure}[h]
\includegraphics[width=6.6cm]{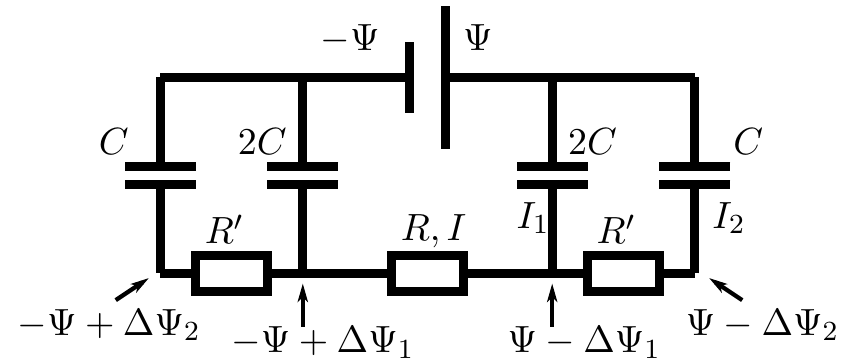}
\caption{Equivalent circuit model for the stack electrode model at $n=2$.}\label{fig_S2}
\end{figure}

For $n=2$, we need to account for the fact that the inner electrode ($i=1$) is facing the electrolyte twice, while the outer electrode ($i=2$) faces it only once: we use $2C$ and $C$ for the different capacitors, respectively,  see \fig{fig_S2}.
For this circuit, the charge on each of the capacitors (index increasing outwards) is $Q_{1}(t)=2C\Delta\Psi_{1}(t)$ and $Q_{2}(t)=C\Delta\Psi_{2}(t)$.
Meanwhile, for the currents $I$, $I_{1}$, and $I_{2}$ we have
$I=I_{1}+I_{2}$ (Kirchoff's law), and $IR=2[\Psi-\Delta\Psi_{1}(t)]$ and $I_{2}R'=\Delta\Psi_{1}(t)-\Delta\Psi_{2}(t)$ (Ohm's law).
Combining the above equations gives two coupled differential equations 
\begin{subequations}\label{coupledODEn2}
\begin{eqnarray}
\Delta \dot{\Psi}_{1}(t)&=&\frac{I_{1}}{2C}=\frac{\Psi}{RC}-\left(\frac{1}{RC}+\frac{1}{2R'C}\right)\Delta\Psi_{1}(t)\nn
&&\quad\quad\quad+\frac{1}{2R'C}\Delta\Psi_{2}(t)\,, \\
\Delta \dot{\Psi}_{2}(t)&=&\frac{I_{2}}{C}=\frac{1}{R'C}\Delta\Psi_{1}(t)-\frac{1}{R'C}\Delta\Psi_{2}(t)\,,\hspace{1.0cm} 
\end{eqnarray}
\end{subequations}
subject to the initial conditions
\begin{subequations}
\begin{eqnarray}
\Delta\Psi_{1}(t=0)&=&0\,,\\
\Delta\Psi_{2}(t=0)&=&0\,.
\end{eqnarray}
\end{subequations}
In matrix form, \eqr{coupledODEn2} reads
\begin{eqnarray}
\dot{
\begin{pmatrix}
\Delta \Psi_{1}(t)\\
\Delta \Psi_{2}(t)\\
\end{pmatrix}}
&=&-\frac{1}{2R'C}
\begin{pmatrix}
1+2R'/R & -1 \\
 -2 &  2 \\
\end{pmatrix}
\begin{pmatrix}
\Delta \Psi_{1}(t)\\
\Delta \Psi_{2}(t)\\
\end{pmatrix}\nn
&&+
\frac{\Psi}{RC}
\begin{pmatrix}
1 \\
 0 \\
\end{pmatrix}
\,,
\end{eqnarray}
which can be written as
$\dot{X}=Y-MX$,
with
\begin{eqnarray}
X\equiv\begin{pmatrix}
\Delta \Psi_{1}(t)\\
\Delta \Psi_{2}(t)\\
\end{pmatrix}
\quad,\quad
Y\equiv\frac{\Psi}{RC}\begin{pmatrix}
1 \\
 0 \\
\end{pmatrix}
\quad,\quad\nn
M\equiv\frac{1}{2R'C}
\begin{pmatrix}
1+2R'/R & -1 \\
 -2 &  2 \\
\end{pmatrix}
\equiv UDU^{-1}\,,
\end{eqnarray}
with $UU^{-1}=U^{-1}U=1$ and $D=\ds{\frac{1}{2R'C}}\begin{pmatrix}
\lambda_{+} & 0 \\
 0 &  \lambda_{-} \\
\end{pmatrix}$.
Here, $U$ is composed of the orthonormal eigenvectors of $M$, with eigenvalues $\lambda_{\pm}$.
To find the formal solution to $\dot{X}=Y-MX$, note that $U^{-1}\dot{X}(t)=U^{-1}Y-DU^{-1}X(t)$ is solved by $U^{-1}X(t)=\exp{[-Dt]}a+D^{-1}U^{-1}Y$.
The constant $a$ is fixed by the boundary condition $U^{-1}X(t=0)=0$,
\begin{eqnarray}
U^{-1}X(0)&=&a+D^{-1}U^{-1}Y=0\nn
\Rightarrow a&=&-D^{-1}U^{-1}Y\,,
\end{eqnarray}
such that the full solution reads $X(t)=U(1-\exp{[-Dt]})D^{-1}U^{-1}Y$, which can be written as
\begin{widetext}
\begin{eqnarray}
X(t)&=&U
\begin{pmatrix}
\displaystyle{\left[1-\exp{\left(-\frac{\lambda_{+} t}{2R'C}\right)}\right]\frac{1}{\lambda_{+}}} & 0 \\
 0 & \displaystyle{\left[1-\exp{\left(-\frac{\lambda_{-} t}{2R'C}\right)}\right]\frac{1}{\lambda_{-}}}  \nonumber\\
\end{pmatrix}
U^{-1}Y\,.
\end{eqnarray}
\end{widetext}
We see that the late-time relaxation of $X(t)$ is determined by the smallest eigenvalue of $M$.
For the stack electrode system, the resistance scales linearly with the width of the respective electrolyte regions.
Specifying to $H=L$, we find $R'/R=H/(2L)=1/2$.
The  eigenvalues of $M$ then follow from $(2-\lambda)(2-\lambda)-2=0$, which amounts to $\lambda_{\pm}=2\pm\sqrt{2}$.
We see that the long-time relaxation is determined by $\tau=2R'C/\lambda_{-}=RC/\lambda_{-}= (2+\sqrt{2})\kappa^{-1}L/D$.

\subsection{General $n$ case}
We redraw \fig{fig1}(c) and introduce additional notation in \fig{fig_S3}. Similar to the previous subsection,
the outer capacitor has a capacitance $C$, while all inner capacitors have $2C$, as they mimic electrodes with electrolyte on either sides.
The charge on each of the capacitors is
\begin{subequations}
\begin{eqnarray}
Q_{1}(t)&=&2C\Delta\Psi_{1}(t)\,,\\
...\nonumber\\
Q_{n-1}(t)&=&2C\Delta\Psi_{n-1}(t)\\
Q_{n}(t)&=&C\Delta\Psi_{n}(t)\,.\label{seq:Qn}
\end{eqnarray}
\end{subequations}
Kirchoff's law states that
\begin{subequations}
\begin{eqnarray}
I&=&I_{1}+I_{12}\,,\\
I_{12}&=&I_{2}+I_{23}\,,\\
...\nonumber\\
I_{(n-2)(n-1)}&=&I_{(n-1)}+I_{(n-1)n}\,,\\
I_{(n-1)n}&=&I_{n}\,,
\end{eqnarray}
\end{subequations}
which gives $I=I_{1}+I_{2}+....+I_{n}$. For the different currents,  Ohm's law reads
\begin{subequations}
\begin{eqnarray}
IR&=&2[\Psi-\Delta\Psi_{1}(t)]\,,\\
I_{12}R'&=&\Delta\Psi_{1}(t)-\Delta\Psi_{2}(t)\,,\\
I_{23}R'&=&\Delta\Psi_{2}(t)-\Delta\Psi_{3}(t)\,,\\
...\nonumber\\
I_{(n-1)n}R'&=&\Delta\Psi_{(n-1)}(t)-\Delta\Psi_{n}(t)\,.
\end{eqnarray}
\end{subequations}
Combining the above equations gives $n$ coupled differential equations
\begin{widetext}

\begin{figure}
\includegraphics[width=10.6cm]{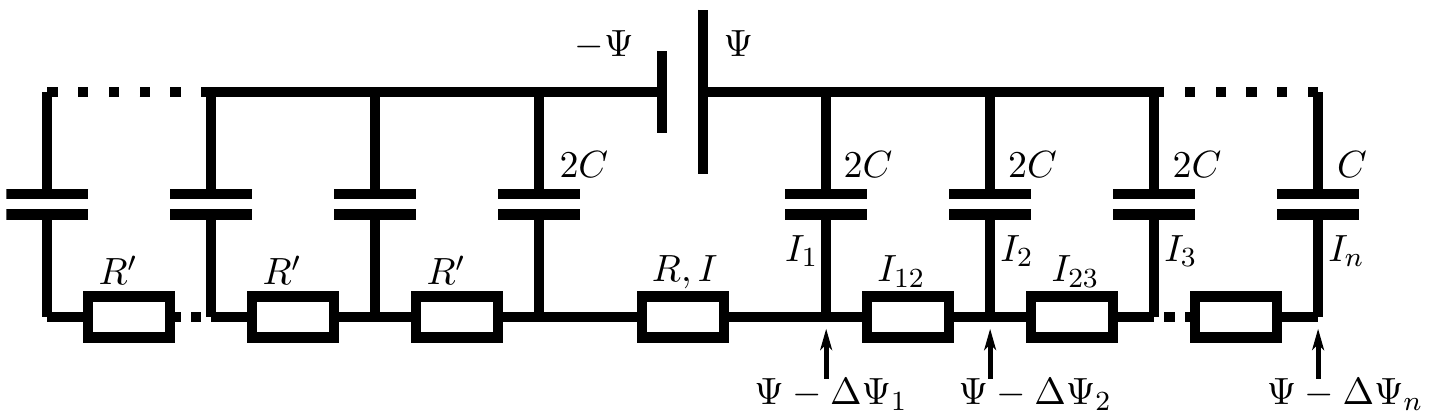}
\caption{Equivalent circuit model for the stack electrode model for general $n$.}\label{fig_S3}
\end{figure}
\begin{subequations}\label{coupledODEn}
\begin{eqnarray}
\Delta \dot{\Psi}_{1}(t)&=&\frac{I_{1}}{2C}=\frac{I-I_{12}}{2C}
=\frac{\Psi}{RC}-\left(\frac{1}{RC}+\frac{1}{2R'C}\right)\Delta\Psi_{1}(t)+\frac{1}{2R'C}\Delta\Psi_{2}(t) \,,\\
\Delta \dot{\Psi}_{2}(t)&=&\frac{I_{2}}{2C}=\frac{I_{12}-I_{23}}{2C}
=\frac{1}{2R'C}\Delta\Psi_{1}(t)-\frac{1}{R'C}\Delta\Psi_{2}(t)+\frac{1}{2R'C}\Delta\Psi_{3}(t) \,,\\
...\nonumber\\
\Delta \dot{\Psi}_{n}(t)&=&\frac{I_{n}}{C}=\frac{1}{R'C}\Delta\Psi_{(n-1)}(t)-\frac{1}{R'C}\Delta\Psi_{n}(t)\,,\label{seq:psin} 
\end{eqnarray}
\end{subequations}
subject to $n$ initial conditions: $\Delta\Psi_{i}(t=0)=0$ for each $i=\{1,..,n\}$.
In matrix form, \eqr{coupledODEn} reads
\begin{eqnarray}\label{eq:matrixn}
\dot{
\begin{pmatrix}
\Delta \Psi_{1}(t)\\
\Delta \Psi_{2}(t)\\
\Delta \Psi_{3}(t)\\
\vdots\\
\Delta \Psi_{n}(t)\\
\end{pmatrix}}
={\frac{\Psi}{RC} }
\begin{pmatrix}
1\\
 0 \\
 .\\
 .\\
 0\\
\end{pmatrix}
-
\frac{1}{2R'C}
\begin{pmatrix}
\ds{1+\frac{H}{L(n-1)}}& -1 & &&&\\
 -1 &  2 & -1&& & \\
  & -1 &  \ddots &\ddots  &&\\
 &&\ddots& \ddots&-1\\
     &  &   &-1 & 2&-1\\
    &  &    & & -2&2\\
\end{pmatrix}
\begin{pmatrix}
\Delta \Psi_{1}(t)\\
\Delta \Psi_{2}(t)\\
\Delta \Psi_{3}(t)\\
 \vdots\\
\Delta \Psi_{n}(t)\\
\end{pmatrix}\,,
\end{eqnarray}
\end{widetext}
which we again write as $\dot{X}=Y-MX$, where $M$ is now a tridiagonal matrix that is Toeplitz except for two elements ($M_{1,1}$ and $M_{n,n-1}$). Note that, for the top left element we used $H=h(n-1)$ to rewrite $R'/R=h/(2L)=H/[2L(n-1)]$ hence $1+2R'/R=1+H/[L(n-1)]$. Again diagonalizing $M$ as $M=UDU^{-1}$, the formal solution to \eqr{eq:matrixn} reads
\begin{eqnarray}\label{eq:generalnsolution}
X(t)&=&U(1-\exp{[-Dt]})D^{-1}U^{-1}Y\,,
\end{eqnarray}
with
\begin{align}\label{eq:generalnsolution2}
&1-\exp{[-Dt]}\nn
&=1-\begin{pmatrix}
\ds{\exp{\left[-\frac{t\lambda_{1}}{2R'C}\right]}}&  & \\
        &  \ddots & \\
        & & \ds{\exp{\left[-\frac{t\lambda_{n}}{2R'C}\right]}}\\
\end{pmatrix}
\end{align}

We used the above model to derive two key results of the main text. First, to obtain the data in \fig{fig2}(b) we diagonalized $M$ numerically for the case $n=5$. With the resulting equivalent circuit potential differences $\Delta \Psi_{i}(t)$ encoded in $X(t)$ [\eqr{eq:generalnsolution}], we determine the surface charge densities $A e\sigma_{i}(t)=C_{i}\Delta \Psi_{i}(t)$ on the electrodes of the stack electrode model. To do that, we use  $C=A \epsilon \kappa$, $\Phi=e\Psi/\kbt$, and $X(t)=\Psi f(t)$, to write $\sigma_{i}(t)=[\kbt\epsilon/ (e^2\kappa^{-1})]\Phi f(t)=2\rhos \kappa^{-1} \Phi f(t) $.
Results are shown in \fig{fig2}(b) with lines. In the same figure, we show numerical results of the PNP equation [Eqs.~\eqref{eq:1}, \eqref{eq:2}, and~\eqref{eq:3}] with symbols. We see that, for $\Phi\ll1$ and $\kappa L>\kappa h\gg1$, the equivalent circuit model describes the behavior of the stack electrode model very well.

\begin{figure}[b]
\includegraphics[width=8.6cm]{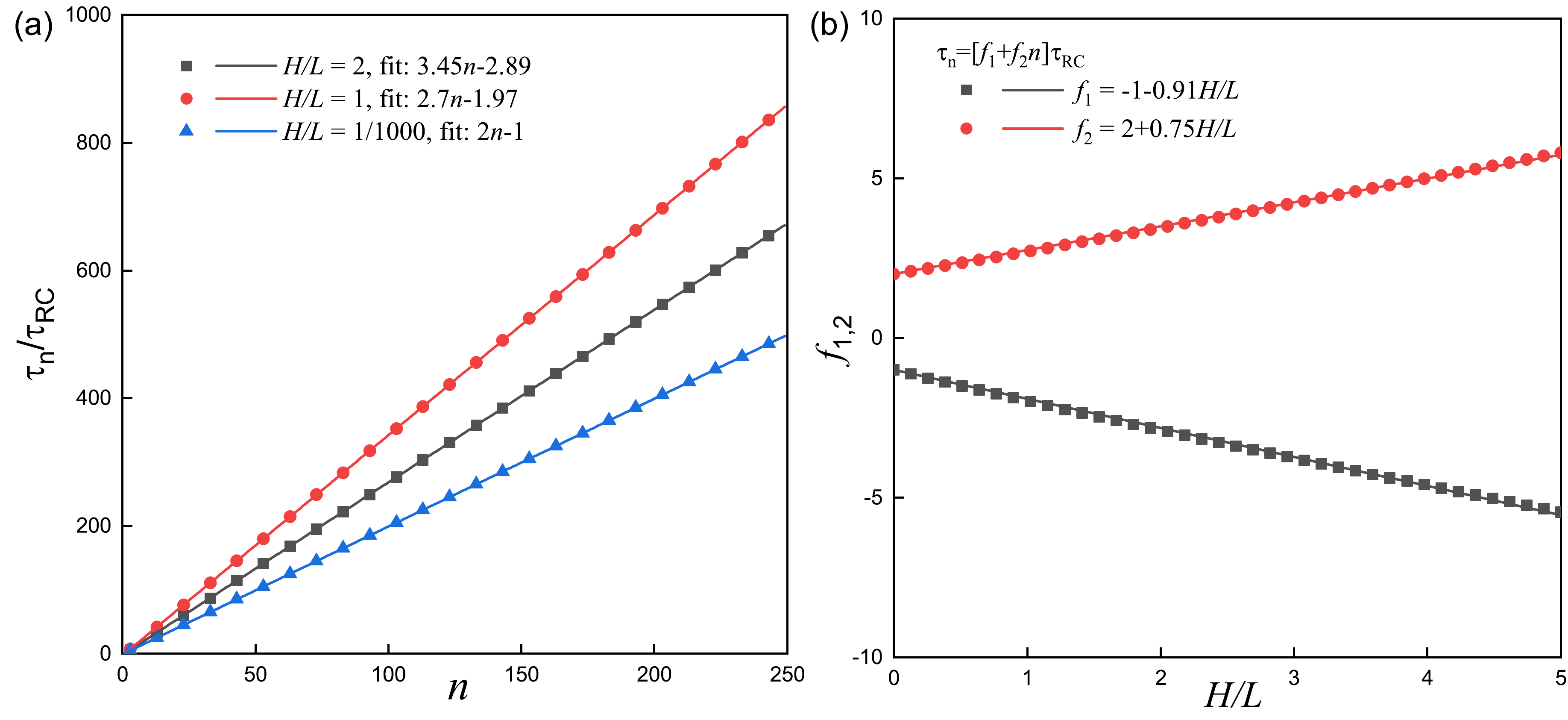}
\caption{ The slowest relaxation time $\tau_{n}$ of a $n$-electrode model as obtained with the smallest eigenvalue $\lambda_{-}$ of the tridiagonal matrix $M$ [cf. \eqr{eq:matrixn}]. The lines in (a) indicate the least-squares fit $\tau_n/\tau_{RC}=f_1+f_2 n$ through the data up to $n=250$. In (b) we then find the fits $ f_1=-1 - 0.91H/L$, and $f_2=2+0.75 H/L$. }
\label{fig_S4}
\end{figure}

Second, to obtain \eqr{eq:4}, we proceed as follows.
Writing $\lambda_{-}=\min\{\lambda_{1},..., \lambda_{n}\}$, the longest relaxation time reads $\tau_{n}=2R'C/\lambda_{-}=\tau_{RC} \times 2H/[L (n-1)\lambda_{-}]$.
We diagonalized $M$ numerically for various $n$ and $H/L$ to find $\lambda_{-}$.
The resulting $\tau_{n}$, shown with symbols in Fig.~\ref{fig_S4}(a), can be accurately approximated by $\tau_{n}=\tau_{RC}(f_{1}+f_{2}n)$: for the three $H/L$ values considered, the residual of a least square fit was at most $R^2=0.00034$.
In Fig.~\ref{fig_S4}(b) we see that the coefficients $f_{1}$ and $f_{2}$ are roughly linear in $H/L$.
We performed a second least-squares fit to determine the coefficients $a_{1}, b_{1}, a_{2}$, and  $b_{2}$ in  $f_{1}=a_{1}+b_{1}H/L$ and $f_{2}=a_{2}+b_{2}H/L$.
Note, however, that we can determine $a_{1}$ and $b_{1}$ analytically.
In the limit $H/L\to0$, i.e., $R'/R\to0$, all capacitors in Fig.~\ref{fig_S3} are connected in parallel, yielding a total capacitance $C_{tot}=(2n-1)C$.
Equation \eqref{seq:nis1} now applies, with $C$ replaced by $C_{tot}$: we find $\tau_{n}=\tau_{RC}(2n-1)$.
This means that we needs to  constrain  $a_{1}=-1$ and $a_{2}=2$ when fitting $f_{1}$ and $f_{2}$.
This ensures that, for $n=1$ (for which $H/L=0$), we correctly reproduce $\tau_{n}=\tau_{RC}$.
Fitting to $f_{1}$ and $f_{2}$-data at $n=250$ [Fig.~\ref{fig_S4}(a)], we find $b_{1}= -0.91$ and  $b_{2}=0.75$  [Fig.~\ref{fig_S4}(b)].
%at second order in H/L and at n=250, the coefficients are [-1.         -0.97995157  0.01854346] [2.         0.70160923 0.0119437 ], and appear in round of form in the plot.
Hence, the equivalent-circuit model predicts that all electrodes relax exponentially at late times with the same
time constant\begin{equation}\label{eq:ana}
\frac{\tau_{n}}{\tau_{RC}}= \left[2+0.75\frac{H}{L}\right]n-1-0.91\frac{H}{L}.
\end{equation}
While the numerical factors in this equation were fitted to data up to $n = 250$, we have performed the same calculation up to $n = 90000$ and found that \eqr{eq:ana} estimated the relaxation time within $1.3 \%$. However, fits to $f_1$ and $f_2$ up to second order in $H/L$ through the data up to $n=250$ yield a corresponding $n=90000$ relaxation time prediction accurate within $0.5\%$.
Interestingly, we see that $\tau_n$ depends on salinity solely through $\tau_{RC}$; hence, $\tau_{n}$ decreases with increasing salt concentration.

\begin{figure}[h]
\includegraphics[width=6.0cm]{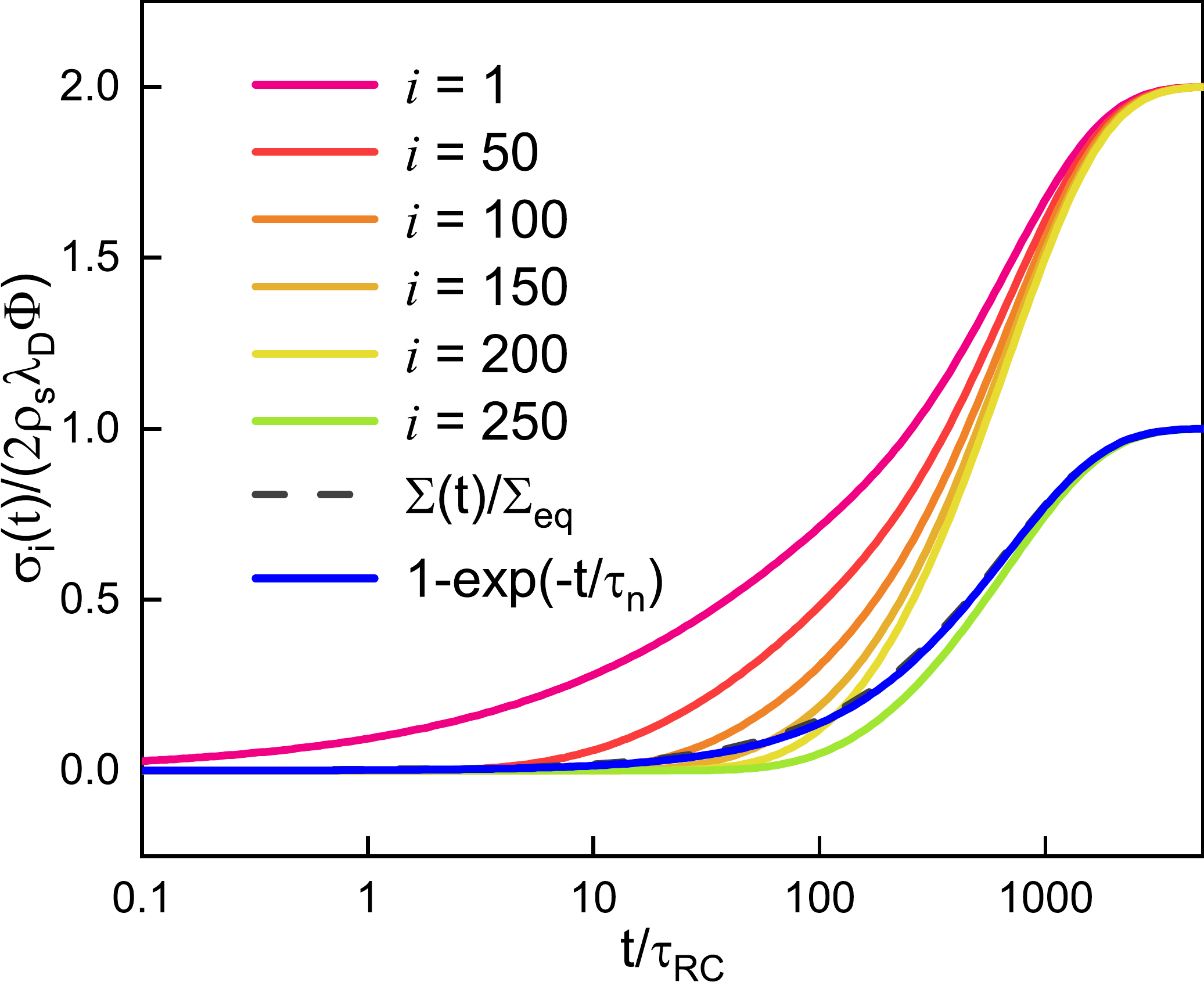}
\caption{Time dependence of the scaled surface charge densities $\sigma_{i}(t)$ for $i=\{1,50,150,200,250\}$  for $H/L=1$, and $n=250$, as predicted by the equivalent circuit model. The colormap goes from red purple ($i=1$) to light green ($i=n$). Shown also are the normalized total surface charge $\Sigma(t)/\Sigma_{eq}$ (dash black) and the function $1-\exp(-t/\tau_n)$ (blue).}
\label{fig_S5}
\end{figure}

Figure \ref{fig_S5} is similar to \fig{fig2}(b), but we now show equivalent circuit predictions for $n=250$ instead of $n=5$. Shown also are the function $1-\exp(-t/\tau_n)$ (blue) and the scaled total surface charge $\Sigma(t)/\Sigma_{eq}$  (red), with  $\Sigma(t)=\sigma_{1}(t)+\sigma_{2}(t)+...+\sigma_n(t)$ and $\Sigma_{eq}=\Sigma(t/\tau_{RC}\to\infty)$. Even though $\Sigma(t)$ harbors $n$ decaying exponential modes [related to all eigenvalues $\lambda_{i}$, see Eqs.~\eqref{eq:generalnsolution} and \eqref{eq:generalnsolution2}], we see that the total surface charge is approximated well by a single exponential decay with the longest relaxation time $\tau_{n}$. In other words, the stack electrode model suggests that $\tau_n$ gives a good estimate for the relaxation of the charge on all $n$ electrodes.

\subsection{Relation to transmission line model}\label{TLappendix}
The circuit shown in \fig{fig_S3} differs from the transmission line model in one place: the capacitance of the $n$-th electrode is $C$ instead of $2C$. Tracing back the steps of the derivation in the previous subsection, we find that for the transmission line model, Eqs.~\eqref{seq:Qn} and \eqref{seq:psin} obtain a factor $2$ and $1/2$, respectively. This yields the same differential equation $\dot{X}=Y-MX$ where $M$ is now given by
\begin{eqnarray}
M=\frac{1}{2R'C}\begin{pmatrix}
\ds{1+\frac{H}{L(n-1)}}& -1 & &&&\\
 -1 &  2 & -1&& & \\
  & -1 &  \ddots &\ddots  &&\\
 &&\ddots& \ddots&-1\\
     &  &   &-1 & 2&-1\\
    &  &    & & -1&1\\
\end{pmatrix}
\hspace{1cm}
\end{eqnarray}
With the same methods as before, one could determine the late-time relaxation timescale, whose $H/L\to0$ limit now reads $\tau_{n}=2n \tau_{RC}$ because $C_{tot}=2C$ in this case.

\setcounter{figure}{0}
\renewcommand{\thefigure}{{B}\arabic{figure}}
\renewcommand{\theHfigure}{B\arabic{figure}}

\section{Additional numerical PNP data}\label{Appendix_B}
In the main text we presented numerical results for the PNP equation \eqref{eq:3} and its initial and boundary conditions [Eqs.~\eqref{eq:1} and~\eqref{eq:2}]  for $H/L=1$ and several $\Phi$, $n$, $\kappa L$. In this section we present additional data for other parameters to reinforce some of the main conclusions of our analysis.

\begin{figure}[b]
\includegraphics[width=8.6cm]{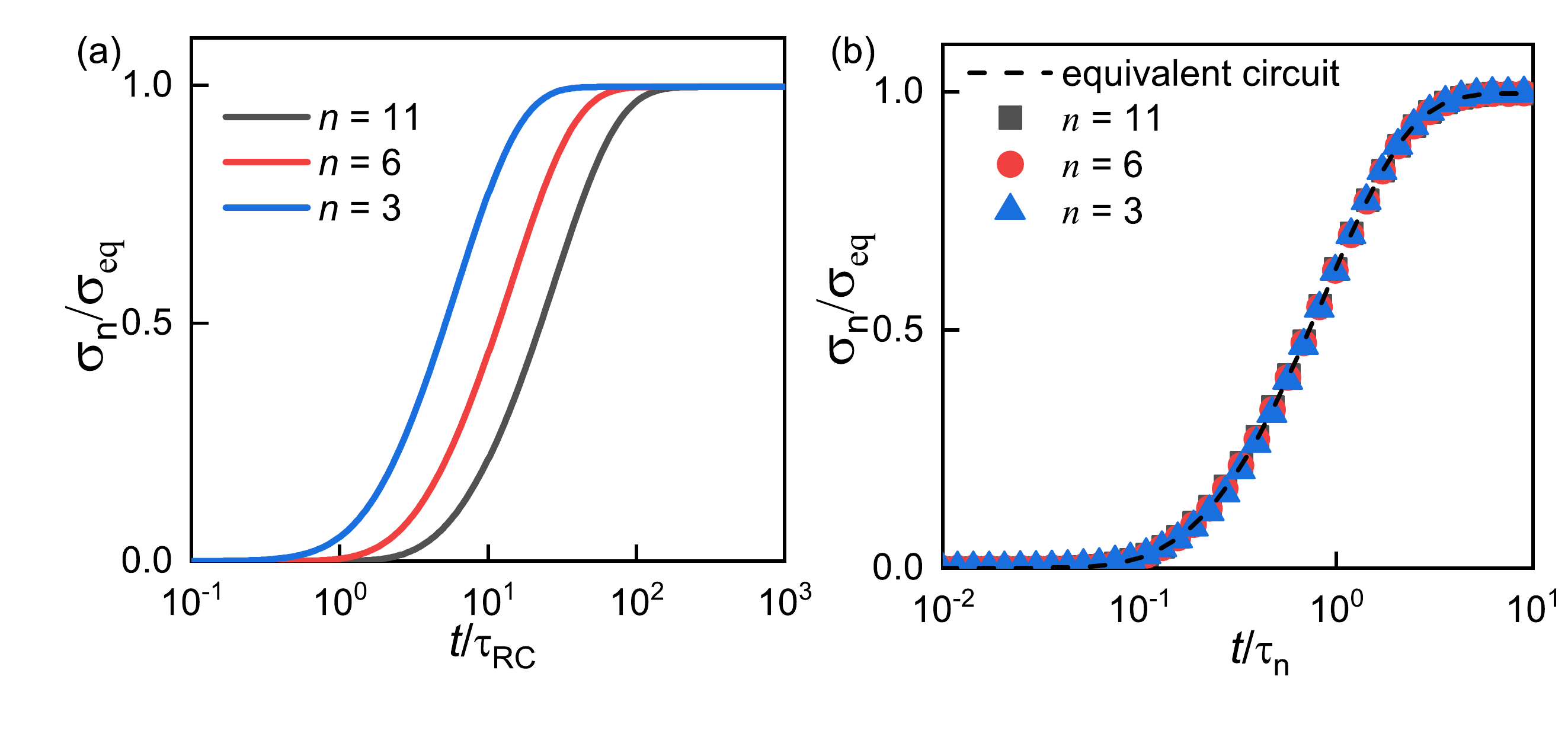}
\caption{The surface charge density $\sigma_{n}(t)$ of the outer electrode from the numerical calculation of Eqs.~\eqref{eq:1}--\eqref{eq:3} for $\Phi=0.001$, $\kappa L= 100$, $H/L=1$ and different $n$. We scale time by either by $\tau_{RC}$ (a) or by $\tau_{n}$ (b). The dashed line is the function $\sigma(t)/\sigma_{eq}=1-\exp[-t/\tau_n]$.}
\label{fig_S6}
\end{figure}

First, we discussed the case $\Phi = 0.001$, $\kappa L = 100$, $H/L=1$, and $n = 5$.
For the same parameters except $n=\{3,6,11\}$, in \fig{fig_S6} we present PNP data for the outer-electrode unit surface charge density $\sigma_{n}(t)$.
Here, time is scaled either by $\tau_{RC}$ [\fig{fig_S6}(a)] or by $\tau_{n}$ [\fig{fig_S6}(b)].
The data collapse that we observe in the latter case again shows the value of the equivalent circuit model in describing the small-$\Phi$ relaxation of our setup.

\begin{figure}
\includegraphics[width=8.6cm]{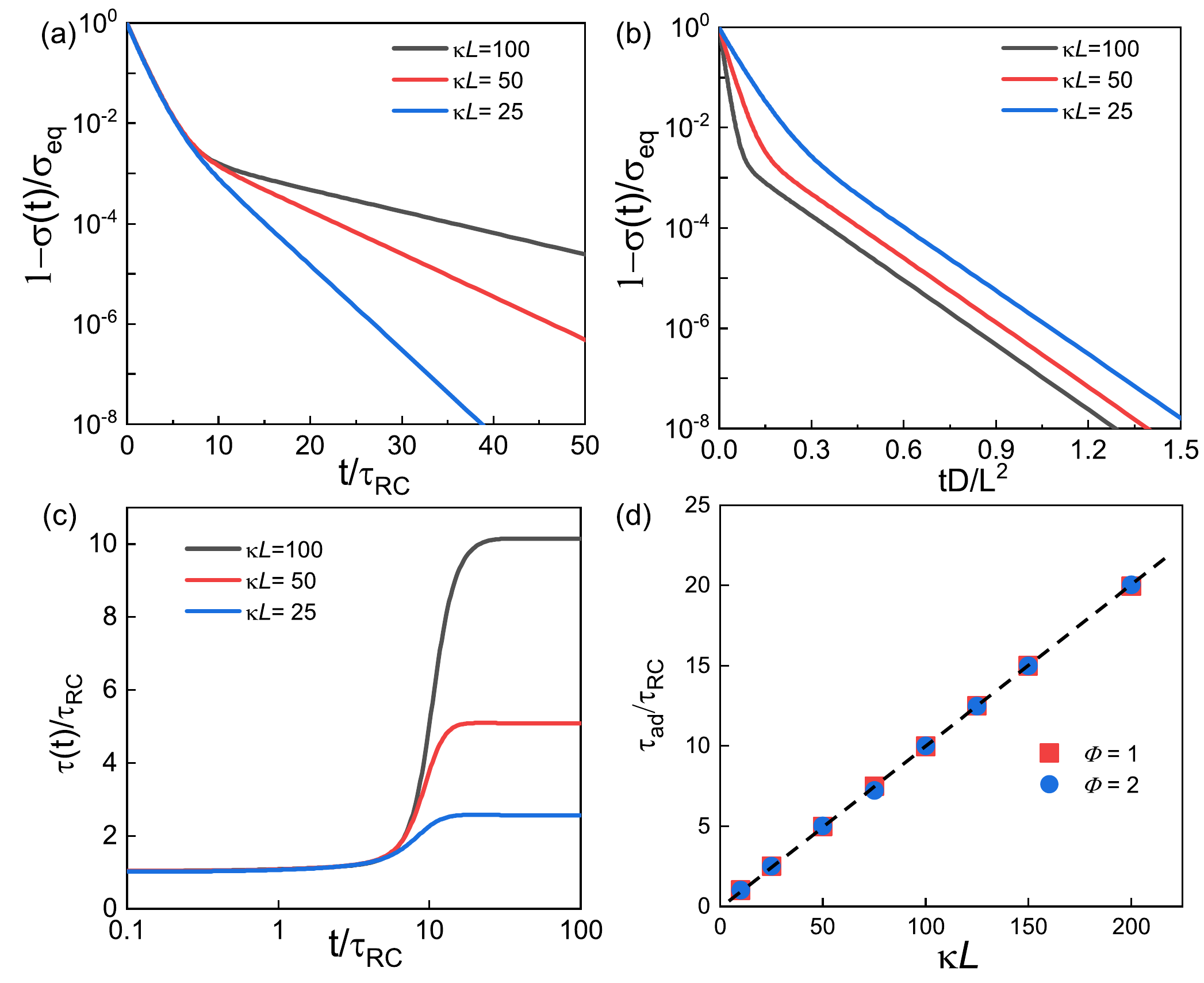}
\caption{(a), (b), and (c) Charging dynamics of the stack-electrode model at $\Phi=1$, $\kappa L=\{25, 50, 100\}$, $H/L=0$ and $n=1$.
We show $1-\sigma(t)/\sigma_{eq}$ with time scaled either by $\tau_{RC}$ (a)  or by $L^2/D$ (b).
(c) The timescale $\tau (t)$ [\eqr{eq:ts}].
(d) The equilibrium relaxation timescale $\tau_{\rm ad}$ as function of $\kappa L$ for $\Phi=1$ and $2$, $n=1$ and $H/L=0$.}
\label{fig_S7}
\end{figure}

Second,  in \fig{fig3} we studied the influence of $\Phi$ and $n$ on our setup through two parameter sets: $\Phi = \{0.01, 0.1 ,1,2\}$,  $\kappa L = 100$, $H/L=1$, and $n = 1$; and $\Phi=2$, $\kappa L=100$, $H/L=1$, and $n=\{2,3,6\}$.
We now use $\Phi = 2$, $\kappa L = \{25, 50, 100\}$, $H/L=0$, and $n = 1$ to study the influence of $\kappa L$.
We show $1-\sigma(t)/\sigma_{eq}$ with time scaled either by $\tau_{RC}$ [\fig{fig_S7}(a)] or by $L^2/D$ [\fig{fig_S7}(b)].
In the first case, $1-\sigma(t)/\sigma_{eq}$ collapses at early times for the different $\kappa L$.
This reinforces our finding of \fig{fig3}(e) where we found a similar early-time collapse for different $\Phi$, from which we concluded that the early-time relaxation of $1-\sigma(t)/\sigma_{eq}$ is described well by the equivalent circuit model.
In the same figure [Fig.~3(e)] we saw that $1-\sigma(t)/\sigma_{eq}$ relaxes on two timescales.
We further investigate the second, slower timescale in \fig{fig_S7}(b), where time is scaled by $L^2/D$.
We see there that $1-\sigma(t)/\sigma_{eq}$ relaxes at late times with the same timescale for the different $\kappa L$.
To characterize both the early- and late-time response, we define a time-dependent function $\tau(t)$,
\begin{equation}\label{eq:ts}
\tau(t)= -\left[\frac{\mathrm{d}\ln(1-\sigma(t)/\sigma_{eq})}{\mathrm{d}t}\right]^{-1}
\end{equation}
that, for a purely exponential charge buildup $\sigma(t)= \sigma_{eq}[1-\exp(-t/\tau^{\star})]$ yields $\tau(t)=\tau^{\star}$.
Figure \ref{fig_S7}(c) presents $\tau (t)$  for the same parameters as used in \fig{fig_S6}(a) and (b).
At early times ($t/\tau_{RC}<1$), $\tau(t)=\tau_{RC}$, while at late times ($t/\tau_{RC}>1$), $\tau(t)$ shoots up to different plateaus $\tau_{\rm ad}$ for different $\kappa L$.
In \fig{fig_S7}(d) we plot $\tau_{\rm ad}$ against $\kappa L$ for $\Phi=1$ and $\Phi=2$ and find a linear relation $\tau_{\rm ad}/ \tau_{RC}=0.1 \kappa L$, hence $\tau_{\rm ad} =  0.1L^2/D$ (if $n=1$).

\begin{figure}
\includegraphics[width=8.6cm]{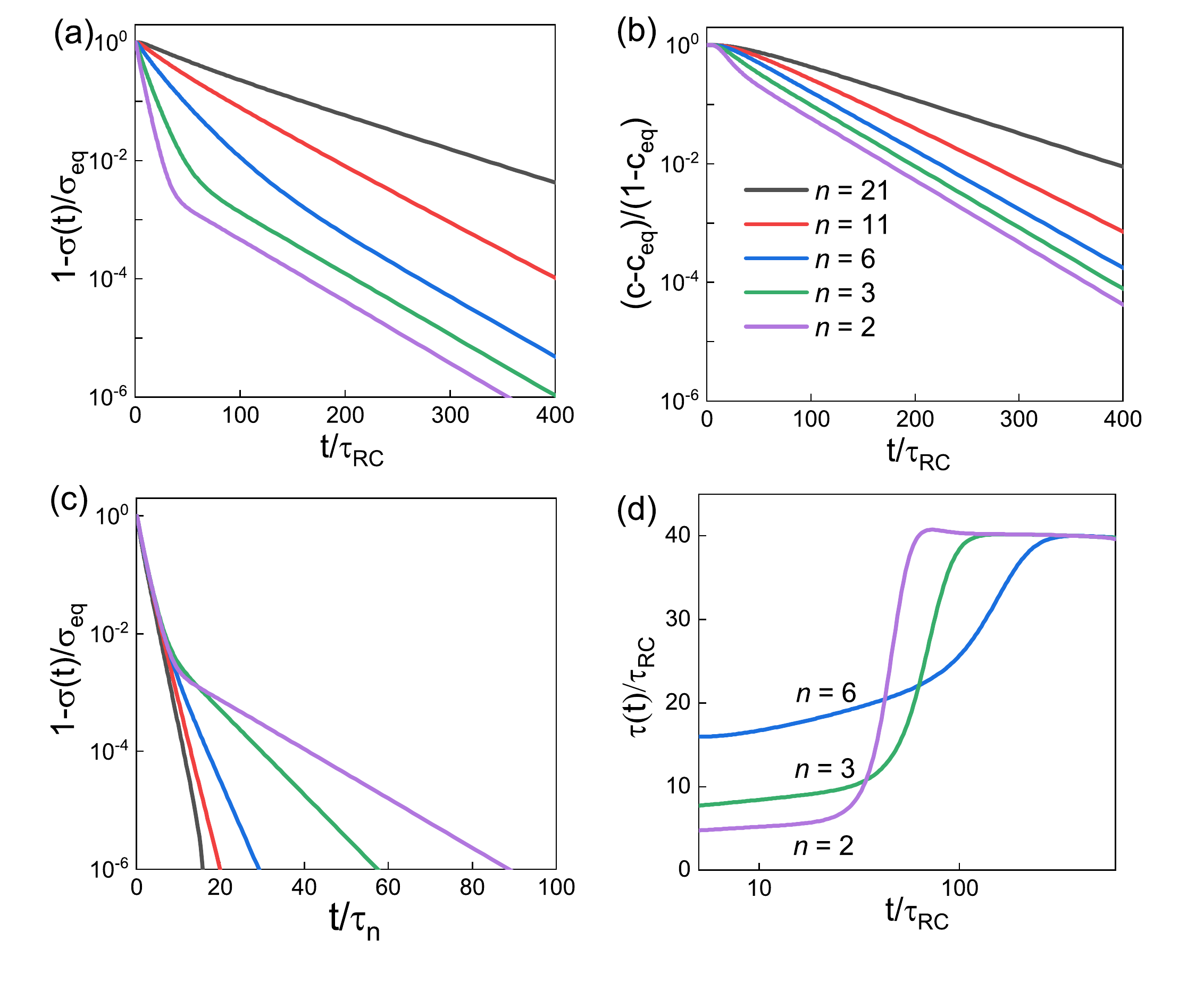}
\caption{(a), (b), and (c) Charging dynamics of the stack-electrode model at $\Phi=2$, $\kappa L=100$, $H/L=1$, and $n=\{2,3,6,11,21 \}$. (a) The charging relaxation $1-\sigma(t)/\sigma_{eq}$ and (b) the concentration decay $(c(t)-c_{eq})/(c(0)-c_{eq})$. In (c) we plot the same data as in (a), but with time scaled by $\tau_{n}$ instead of by $\tau_{RC}$. (d) The $\tau(t)$ [\eqr{eq:ts}] for $n=\{2,3,6\}$ and $\Phi$, $\kappa L$, and $H/L$ as in (a), (b), and (c). }
\label{fig_S8}
\end{figure}

Third, as mentioned above, we discussed the case $\Phi=2$, $\kappa L=100$, $H/L=1$, and $n=\{2,3,6\}$.
Here, we present additional data for the same $\Phi$, $\kappa L$, and $H/L$ for $n=\{2,3,6,11,21 \}$.
Notably, these values correspond to $\kappa h=\{100,50,20,10,5\}$, which means that we probe the behavior of the stack-electrode model for both non-overlapping and overlapping EDLs.
As noted in the main text, \fig{fig_S8}(a) again shows that increased porosity (higher $n$) leads to slower surface charge buildup.
Moreover, we see that  $1-\sigma(t)/\sigma_{eq}$ relaxes at late times with the same timescale for the $n = \{2,3,6\}$, while for $n=11$ and $n=21$ (corresponding to $\kappa h =10$ and $\kappa h =5$) it relaxes slower, arguably because EDLs overlap.
The same conclusions hold for $(c(t)-c_{eq})/(c(0)-c_{eq})$, which we plot in \fig{fig_S8}(b).
In \fig{fig_S8}(c) we plot the same data as in \fig{fig_S8}(a), but with time scaled by $\tau_{n}$ instead of by $\tau_{RC}$.
We see there that $1-\sigma(t)/\sigma_{eq}$  collapses at early times for all $n$ considered, which again confirms the early-time $\tau_{n}$ scaling.
For completeness, in \fig{fig_S8}(d), we again evaluated $\tau(t)$ as defined in \eqr{eq:ts}: similar to the $n=1$ case, two plateaus appear.

\setcounter{figure}{0}
\renewcommand{\thefigure}{{C}\arabic{figure}}
\renewcommand{\theHfigure}{C\arabic{figure}}
\section{Dynamic density functional theory for concentrated electrolytes}\label{Appendix_C}
In the main text, we discussed the charging dynamics of the electrodes by both the equivalent circuit model and the classical Poisson-Nernst-Planck, which both are meaningful for low charging potentials and dilute electrolytes. We now examine higher applied potentials and high concentrations by classical dynamic density functional theory (DDFT), where both the equivalent circuit model and the Poisson-Nernst-Planck equations are expected to fail. For ion diffusion between two charged electrodes, DDFT asserts that the time evolution for the local density profiles of ionic species \cite{Lian2016JCP,jiang2014kinetic}, again follows \eqr{eq:3} with $j_{\pm}(x,t)$ replaced by
	\begin{align}\label{eq:ddft}
	j_{\pm}(x,t)&= -\frac{D}{\kbt}\rho_{\pm}(x,t) \partial_{x} \left(\frac{\delta F[\rho_{+}, \rho_{-}]}{\delta \rho_{\pm}(x,t)}\right) ,
	\end{align}
with $F[\rho_{+}, \rho_{-}]$ being the free energy functional, for which a suitable approximation needs to be made.
To model electrolyte solutions, $F[\rho_{+}, \rho_{-}]$ should contain at least an ideal-solution term and the mean electrostatic potential. 
Moreover, we model contributions from ionic excluded volume effects through the functional described in Ref.~\cite{Lian2016JCP,jiang2014kinetic}. The restricted primitive model is used to represent a room-temperature ionic liquid wherein both cations and anions are approximated by monovalent charged hard spheres. The diameters of cation and anion are both $a=0.5 ~\si{\nano\meter}$  (corresponding to the averaged size of EMI and TFSI), the concentration is $\rho_b=1$ M, the diffusion coefficient for the ionic liquid are  $D=4.3\times10^{-11}~\si{\meter\squared\per\second}$, and the dielectric constant is set as $\varepsilon=2$.

\begin{figure}
	\includegraphics[width=6cm]{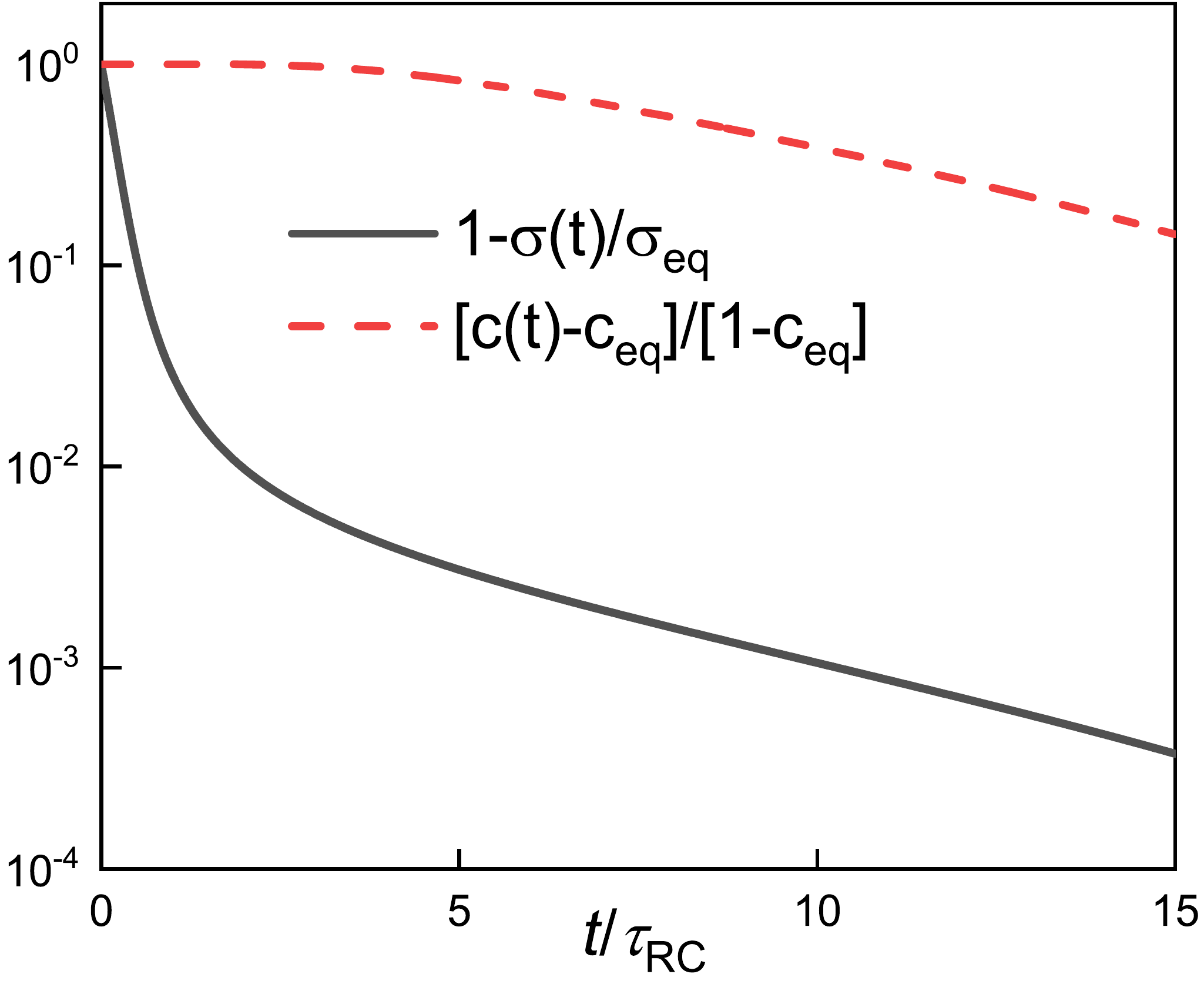}
	\caption{The charge relaxation $1-\sigma(t)/\sigma_{eq}$ (solid line) and the concentration decay $(c(t)-c_{eq})/(1-c_{eq})$ (dashed line) for a 1 M ionic liquid electrolyte at $\Phi=20$, $L/a=12$, $H/L=0$, $\kappa^{-1}=4.86\times10^{-11}~\si{\nano\meter}$, and $n=1$. }
	\label{fig_S10}
\end{figure}

\fig{fig_S10} shows DDFT result for the evolving surface charge density $1-\sigma(t)/\sigma_{eq}$ (solid lines) and the normalized salt concentration at the cell center  $(c(t)-c_{eq})/(1-c_{eq})$ (dashed lines). We observe interesting similarities to the PNP data presented in \fig{fig3}(e): the surface charge density exhibits two distinct relaxation times and, at late times, $1-\sigma(t)/\sigma_{eq}$ and $(c(t)-c_{eq})/(1-c_{eq})$ decay with the same time constant. 
Qualitatively, this is very similar to our PNP results, and illustrate the robustness of our PNP findings regarding the existence of two timescales that govern coupled charge and salt transport. 
However, a direct comparison between our DDFT and PNP results is difficult for the following reasons: First, PNP reduces to Poisson-Boltzmann theory at equilibrium. Boltzmann weights $\exp(\Phi)$ would predict unphysically large counterion densities close to the electrode surfaces.
Second, the RTIL has a tiny Debye length. Recent experimental studies, however, have suggested that, instead of $\kappa^{-1}$, an anomalously large electrostatic screening length $\lambda_S$ governs the electrostatic interactions of dense electrolytes and RTILs \cite{susan2017}. It would be interesting to find whether $\lambda_S$ replaces $\lambda_D$ in the $RC$ time of the $n=1$ circuit. However, as dense electrolytes and RTILs, even at equilibrium, are currently a hotly debated topic, a full characterization of the out-of-equilibrium behavior of our $n>1$ stack electrode model is beyond the scope of the current article. A more detailed and quantitative comparison of the PNP and DDFT results will be presented elsewhere.

\setcounter{figure}{0}
\renewcommand{\thefigure}{{D}\arabic{figure}}
\renewcommand{\theHfigure}{D\arabic{figure}}

\section{Comparison to experimental surface charge build up}\label{Appendix_D}
In the main text, we determined $\tau_n$ and $\tau_{\rm ad}$ for parameters corresponding to the experimental setup of \cit{janssen2017coulometry}.
Here, We reanalyze data of that paper for the decaying electric current $I(t)$ caused by  an applied potential $\Psi=0.3$. Repeated measurements were performed at different $\rho_b$: twice at $\rho_b=\{0.001, 0.01, 0.1\}$ M and four times at $\rho_b=1 $ M.
\begin{figure}[b]
\includegraphics[width=8.6cm]{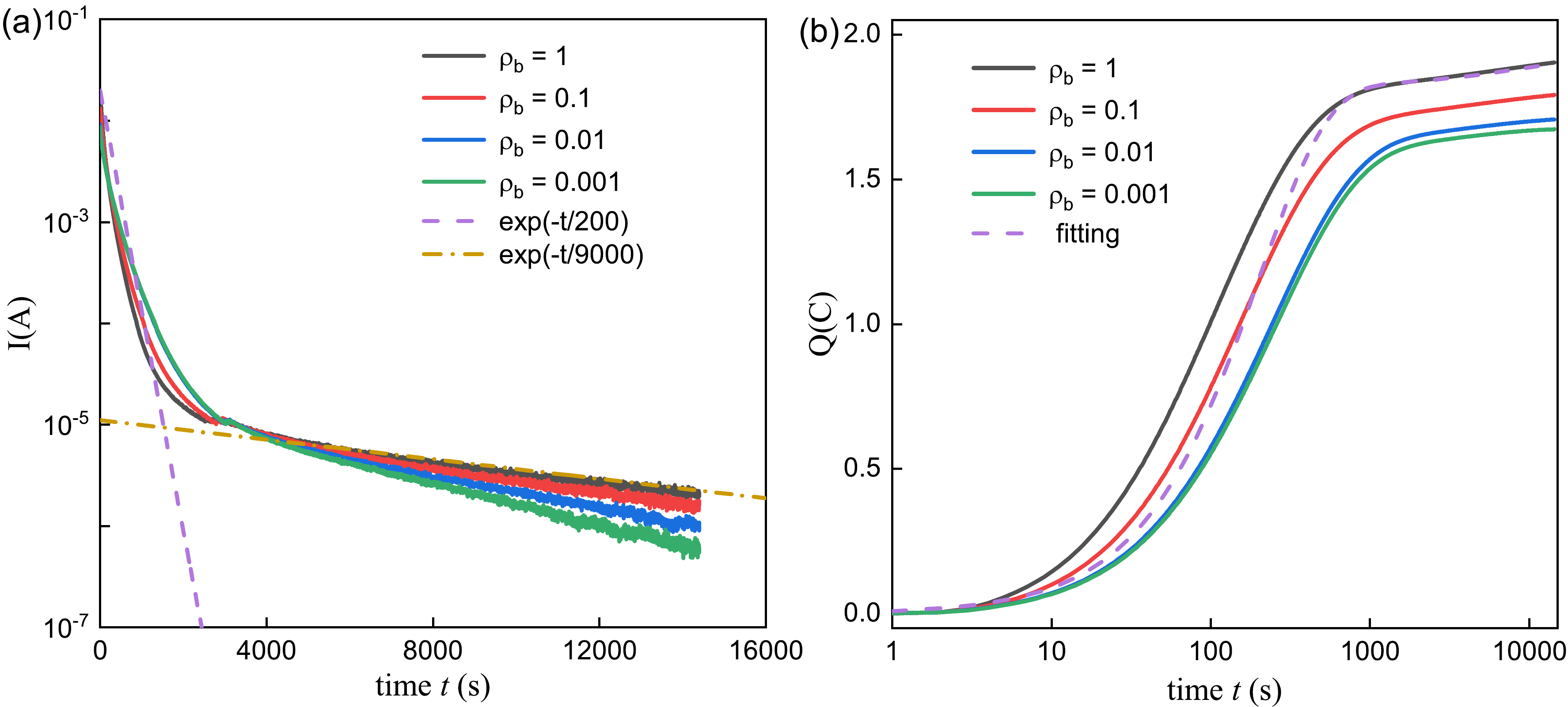}
\caption{(a) Experimental data  of \cit{janssen2017coulometry} (dots) for the current relaxation of porous electrodes subject to a suddenly imposed potential $\Psi=0.3~\si{\volt}$. 
(b) The total surface charge $Q$, found by integrating the data from (a). The dash line is a fit by eye: \mbox{$Q = \{1.8[1- \exp(-t/200)]+0.1[1- \exp(-t/9000)]\}$ C}, whose two time constants were inspired by the dotted and dash-dotted lines in (a).}
\label{fig_S9}
\end{figure}
\mbox{Figure~\ref{fig_S9}(a)} shows the average currents $I(t)$ of those measurements for each $\rho_b $. 
Dotted and dash-dotted lines in \fig{fig_S9}(a) show two exponential decays with time constants $\tau_1=200~\si{\second}$ and $\tau_2=9000~\si{\second}$. With these lines one sees for $\rho_b=1$ M that $I(t)$ decays exponentially at late times, while at early times its behavior is more complicated.
\mbox{Figure~\ref{fig_S9}(b)} shows $Q(t)=\int_{0}^{t}I(t)$, determined numerically by integrating an interpolation through the data of \fig{fig_S9}(a). Also shown with a dashed line is the function $Q (t) = \{1.8 [1-\exp(-t/\tau_{1})]+0.1[1- \exp(-t/\tau_{2})]\}$ C, where the numerical prefactors were fixed by eye. This dashed line describes $Q(t)$ properly after $t\approx 3000~\si{\second}$, underestimating $Q(t)$ at earlier times. 
We saw in \fig{fig_S5} that the stack electrode model displays a similar feature: at intermediate times, the total surface charge $\Sigma(t)$ is larger than $1-\exp(-t/\tau_n)$. The difference between these two observables, however, is much smaller in the stack electrode model than in the experiments. Hence, our stack electrode model does not yet fully explain the experimental data of  \cit{janssen2017coulometry}.
Still, comparing the fitted time constants $\tau_{1}=200~\si{\second}$ and $\tau_{2}=9000~\si{\second}$ to $\tau_n=2.9\times 10^2~\si{\second}$ and $\tau_{\rm ad}=4.8\times 10^2~\si{\second}$ as stated in the main text, respectively, we see that our model predicts both timescales within approximately one order of magnitude. Even though we seem to predict $\tau_{1}$ better than $\tau_{2}$, using a smaller diffusion constant $D=2\times10^{-10}~\si{\meter\squared\per\second}$, to account for slow diffusion in pores \cite{PhysRevE.76.011501, ALLAHYAROV201773}, leads to $\tau_n=2.3\times 10^3~\si{\second}$ and $\tau_{\rm ad}=3.6\times 10^3~\si{\second}$ and predictions for $\tau_{2}$ are better than for $\tau_1$.
\end{appendix}


\begin{thebibliography}{40}%
\makeatletter
\providecommand \@ifxundefined [1]{%
 \@ifx{#1\undefined}
}%
\providecommand \@ifnum [1]{%
 \ifnum #1\expandafter \@firstoftwo
 \else \expandafter \@secondoftwo
 \fi
}%
\providecommand \@ifx [1]{%
 \ifx #1\expandafter \@firstoftwo
 \else \expandafter \@secondoftwo
 \fi
}%
\providecommand \natexlab [1]{#1}%
\providecommand \enquote  [1]{``#1''}%
\providecommand \bibnamefont  [1]{#1}%
\providecommand \bibfnamefont [1]{#1}%
\providecommand \citenamefont [1]{#1}%
\providecommand \href@noop [0]{\@secondoftwo}%
\providecommand \href [0]{\begingroup \@sanitize@url \@href}%
\providecommand \@href[1]{\@@startlink{#1}\@@href}%
\providecommand \@@href[1]{\endgroup#1\@@endlink}%
\providecommand \@sanitize@url [0]{\catcode `\\12\catcode `\$12\catcode
  `\&12\catcode `\#12\catcode `\^12\catcode `\_12\catcode `\%12\relax}%
\providecommand \@@startlink[1]{}%
\providecommand \@@endlink[0]{}%
\providecommand \url  [0]{\begingroup\@sanitize@url \@url }%
\providecommand \@url [1]{\endgroup\@href {#1}{\urlprefix }}%
\providecommand \urlprefix  [0]{URL }%
\providecommand \Eprint [0]{\href }%
\providecommand \doibase [0]{http://dx.doi.org/}%
\providecommand \selectlanguage [0]{\@gobble}%
\providecommand \bibinfo  [0]{\@secondoftwo}%
\providecommand \bibfield  [0]{\@secondoftwo}%
\providecommand \translation [1]{[#1]}%
\providecommand \BibitemOpen [0]{}%
\providecommand \bibitemStop [0]{}%
\providecommand \bibitemNoStop [0]{.\EOS\space}%
\providecommand \EOS [0]{\spacefactor3000\relax}%
\providecommand \BibitemShut  [1]{\csname bibitem#1\endcsname}%
\let\auto@bib@innerbib\@empty
%</preamble>
\bibitem [{\citenamefont {Chmiola}\ \emph {et~al.}(2006)\citenamefont
  {Chmiola}, \citenamefont {Yushin}, \citenamefont {Gogotsi}, \citenamefont
  {Portet}, \citenamefont {Simon},\ and\ \citenamefont
  {Taberna}}]{chmiola2006sci}%
  \BibitemOpen
  \bibfield  {author} {\bibinfo {author} {\bibfnamefont {J.}~\bibnamefont
  {Chmiola}}, \bibinfo {author} {\bibfnamefont {G.}~\bibnamefont {Yushin}},
  \bibinfo {author} {\bibfnamefont {Y.}~\bibnamefont {Gogotsi}}, \bibinfo
  {author} {\bibfnamefont {C.}~\bibnamefont {Portet}}, \bibinfo {author}
  {\bibfnamefont {P.}~\bibnamefont {Simon}},\ and\ \bibinfo {author}
  {\bibfnamefont {P.-L.}\ \bibnamefont {Taberna}},\ }\href
  {https://science.sciencemag.org/content/313/5794/1760} {\bibfield  {journal}
  {\bibinfo  {journal} {Science}\ }\textbf {\bibinfo {volume} {313}},\ \bibinfo
  {pages} {1760} (\bibinfo {year} {2006})}\BibitemShut {NoStop}%
\bibitem [{\citenamefont {Limmer}\ \emph {et~al.}(2013)\citenamefont {Limmer},
  \citenamefont {Merlet}, \citenamefont {Salanne}, \citenamefont {Chandler},
  \citenamefont {Madden}, \citenamefont {van Roij},\ and\ \citenamefont
  {Rotenberg}}]{limmer2013charge}%
  \BibitemOpen
  \bibfield  {author} {\bibinfo {author} {\bibfnamefont {D.~T.}\ \bibnamefont
  {Limmer}}, \bibinfo {author} {\bibfnamefont {C.}~\bibnamefont {Merlet}},
  \bibinfo {author} {\bibfnamefont {M.}~\bibnamefont {Salanne}}, \bibinfo
  {author} {\bibfnamefont {D.}~\bibnamefont {Chandler}}, \bibinfo {author}
  {\bibfnamefont {P.~A.}\ \bibnamefont {Madden}}, \bibinfo {author}
  {\bibfnamefont {R.}~\bibnamefont {van Roij}},\ and\ \bibinfo {author}
  {\bibfnamefont {B.}~\bibnamefont {Rotenberg}},\ }\href
  {https://link.aps.org/doi/10.1103/PhysRevLett.111.106102} {\bibfield
  {journal} {\bibinfo  {journal} {Phys. Rev. Lett.}\ }\textbf {\bibinfo
  {volume} {111}},\ \bibinfo {pages} {106102} (\bibinfo {year}
  {2013})}\BibitemShut {NoStop}%  
\bibitem [{\citenamefont {Forse}\ \emph {et~al.}(2017)\citenamefont {Forse},
  \citenamefont {Griffin}, \citenamefont {Merlet}, \citenamefont
  {Carretero-Gonzalez}, \citenamefont {Raji}, \citenamefont {Trease},\ and\
  \citenamefont {Grey}}]{forse2017direct}%
  \BibitemOpen
  \bibfield  {author} {\bibinfo {author} {\bibfnamefont {A.~C.}\ \bibnamefont
  {Forse}}, \bibinfo {author} {\bibfnamefont {J.~M.}\ \bibnamefont {Griffin}},
  \bibinfo {author} {\bibfnamefont {C.}~\bibnamefont {Merlet}}, \bibinfo
  {author} {\bibfnamefont {J.}~\bibnamefont {Carretero-Gonzalez}}, \bibinfo
  {author} {\bibfnamefont {A.-R.~O.}\ \bibnamefont {Raji}}, \bibinfo {author}
  {\bibfnamefont {N.~M.}\ \bibnamefont {Trease}},\ and\ \bibinfo {author}
  {\bibfnamefont {C.~P.}\ \bibnamefont {Grey}},\ }\href
  {https://www.nature.com/articles/nenergy2016216} {\bibfield  {journal}
  {\bibinfo  {journal} {Nat. Energy}\ }\textbf {\bibinfo {volume} {2}},\
  \bibinfo {pages} {16216} (\bibinfo {year} {2017})}\BibitemShut {NoStop}%
\bibitem [{\citenamefont {Prehal}\ \emph {et~al.}(2018)\citenamefont {Prehal},
  \citenamefont {Koczwara}, \citenamefont {Amenitsch}, \citenamefont
  {Presser},\ and\ \citenamefont {Paris}}]{prehal2018salt}%
  \BibitemOpen
  \bibfield  {author} {\bibinfo {author} {\bibfnamefont {C.}~\bibnamefont
  {Prehal}}, \bibinfo {author} {\bibfnamefont {C.}~\bibnamefont {Koczwara}},
  \bibinfo {author} {\bibfnamefont {H.}~\bibnamefont {Amenitsch}}, \bibinfo
  {author} {\bibfnamefont {V.}~\bibnamefont {Presser}},\ and\ \bibinfo
  {author} {\bibfnamefont {O.}~\bibnamefont {Paris}},\ }\href
  {https://www.nature.com/articles/s41467-018-06612-4} {\bibfield  {journal}
  {\bibinfo  {journal} {Nat. Commun.}\ }\textbf {\bibinfo {volume} {9}},\
  \bibinfo {pages} {4145} (\bibinfo {year} {2018})}\BibitemShut {NoStop}%
\bibitem [{\citenamefont {Simon}\ and\ \citenamefont
  {Gogotsi}(2013)}]{simon2012capacitive}%
  \BibitemOpen
  \bibfield  {author} {\bibinfo {author} {\bibfnamefont {P.}~\bibnamefont
  {Simon}}\ and\ \bibinfo {author} {\bibfnamefont {Y.}~\bibnamefont
  {Gogotsi}},\ }\href {\doibase 10.1021/ar200306b} {\bibfield  {journal}
  {\bibinfo  {journal} {Acc. Chem. Res.}\ }\textbf {\bibinfo {volume} {46}},\
  \bibinfo {pages} {1094} (\bibinfo {year} {2013})}\BibitemShut {NoStop}%
 \bibitem [{\citenamefont {Cheng}\ \emph {et~al.}(2016)\citenamefont {Cheng},
  \citenamefont {Jiang}, \citenamefont {Garvey}, \citenamefont {Wang},
  \citenamefont {Simon}, \citenamefont {Liu},\ and\ \citenamefont
  {Li}}]{cheng2016ion}%
  \BibitemOpen
  \bibfield  {author} {\bibinfo {author} {\bibfnamefont {C.}~\bibnamefont
  {Cheng}}, \bibinfo {author} {\bibfnamefont {G.}~\bibnamefont {Jiang}},
  \bibinfo {author} {\bibfnamefont {C.~J.}\ \bibnamefont {Garvey}}, \bibinfo
  {author} {\bibfnamefont {Y.}~\bibnamefont {Wang}}, \bibinfo {author}
  {\bibfnamefont {G.~P.}\ \bibnamefont {Simon}}, \bibinfo {author}
  {\bibfnamefont {J.~Z.}\ \bibnamefont {Liu}},\ and\ \bibinfo {author}
  {\bibfnamefont {D.}~\bibnamefont {Li}},\ }\href
  {https://advances.sciencemag.org/content/2/2/e1501272} {\bibfield  {journal}
  {\bibinfo  {journal} {Sci. Adv.}\ }\textbf {\bibinfo {volume} {2}},\ \bibinfo
  {pages} {e1501272} (\bibinfo {year} {2016})}\BibitemShut {NoStop}% 
\bibitem [{\citenamefont {Mouterde}\ \emph {et~al.}(2019)\citenamefont
  {Mouterde}, \citenamefont {Keerthi}, \citenamefont {Poggioli}, \citenamefont
  {Dar}, \citenamefont {Siria}, \citenamefont {Geim}, \citenamefont {Bocquet},\
  and\ \citenamefont {Radha}}]{mouterde2019molecular}%
  \BibitemOpen
  \bibfield  {author} {\bibinfo {author} {\bibfnamefont {T.}~\bibnamefont
  {Mouterde}}, \bibinfo {author} {\bibfnamefont {A.}~\bibnamefont {Keerthi}},
  \bibinfo {author} {\bibfnamefont {A.~R.}~\bibnamefont {Poggioli}}, \bibinfo
  {author} {\bibfnamefont {S.~A.}~\bibnamefont {Dar}}, \bibinfo {author}
  {\bibfnamefont {A.}~\bibnamefont {Siria}}, \bibinfo {author} {\bibfnamefont
  {A.~K.}~\bibnamefont {Geim}}, \bibinfo {author} {\bibfnamefont {L.}~\bibnamefont
  {Bocquet}},\ and\ \bibinfo {author} {\bibfnamefont {B.}~\bibnamefont
  {Radha}},\ }\href {https://www.nature.com/articles/s41586-019-0961-5}
  {\bibfield  {journal} {\bibinfo  {journal} {Nature}\ }\textbf {\bibinfo
  {volume} {567}},\ \bibinfo {pages} {87} (\bibinfo {year} {2019})}\BibitemShut
  {NoStop}%
\bibitem [{\citenamefont {de~Levie}(1963)}]{DELEVIE1963751}%
  \BibitemOpen
  \bibfield  {author} {\bibinfo {author} {\bibfnamefont {R.}~\bibnamefont
  {de~Levie}},\ }\href {\doibase https://doi.org/10.1016/0013-4686(63)80042-0}
  {\bibfield  {journal} {\bibinfo  {journal} {Electrochim. Acta}\ }\textbf
  {\bibinfo {volume} {8}},\ \bibinfo {pages} {751 } (\bibinfo {year}
  {1963})}\BibitemShut {NoStop}%
\bibitem [{\citenamefont {Mirzadeh}\ \emph {et~al.}(2014)\citenamefont
  {Mirzadeh}, \citenamefont {Gibou},\ and\ \citenamefont
  {Squires}}]{PhysRevLett.113.097701}%
  \BibitemOpen
  \bibfield  {author} {\bibinfo {author} {\bibfnamefont {M.}~\bibnamefont
  {Mirzadeh}}, \bibinfo {author} {\bibfnamefont {F.}~\bibnamefont {Gibou}},\ and\ \bibinfo {author} {\bibfnamefont {T.~M.}\ \bibnamefont {Squires}},\
  }\href {\doibase 10.1103/PhysRevLett.113.097701} {\bibfield  {journal}
  {\bibinfo  {journal} {Phys. Rev. Lett.}\ }\textbf {\bibinfo {volume} {113}},\
  \bibinfo {pages} {097701} (\bibinfo {year} {2014})}\BibitemShut {NoStop}%
\bibitem [{\citenamefont {Tivony}\ \emph {et~al.}(2018)\citenamefont {Tivony},
  \citenamefont {Safran}, \citenamefont {Pincus}, \citenamefont {Silbert},\
  and\ \citenamefont {Klein}}]{tivony2018charging}%
  \BibitemOpen
  \bibfield  {author} {\bibinfo {author} {\bibfnamefont {R.}~\bibnamefont
  {Tivony}}, \bibinfo {author} {\bibfnamefont {S.}~\bibnamefont {Safran}},
  \bibinfo {author} {\bibfnamefont {P.}~\bibnamefont {Pincus}}, \bibinfo
  {author} {\bibfnamefont {G.}~\bibnamefont {Silbert}},\ and\ \bibinfo
  {author} {\bibfnamefont {J.}~\bibnamefont {Klein}},\ }\href
  {https://www.nature.com/articles/s41467-018-06364-1} {\bibfield  {journal}
  {\bibinfo  {journal} {Nat. Commun.}\ }\textbf {\bibinfo {volume} {9}},\
  \bibinfo {pages} {4203} (\bibinfo {year} {2018})}\BibitemShut {NoStop}%
\bibitem [{\citenamefont {Helseth}(2019)}]{HELSETH2019}%
\BibitemOpen
\bibfield  {author} {\bibinfo {author} {\bibfnamefont {L.}~\bibnamefont
		{Helseth}},\ }\href {\doibase https://doi.org/10.1016/j.est.2019.100912}
{\bibfield  {journal} {\bibinfo  {journal} {J. Energy Storage}\ }\textbf
	{\bibinfo {volume} {25}},\ \bibinfo {pages} {100912} (\bibinfo {year}
	{2019})}\BibitemShut {NoStop}%  
\bibitem [{\citenamefont {Feng}\ and\ \citenamefont
  {Cummings}(2011)}]{feng2011supercapacitor}%
  \BibitemOpen
  \bibfield  {author} {\bibinfo {author} {\bibfnamefont {G.}~\bibnamefont
  {Feng}}\ and\ \bibinfo {author} {\bibfnamefont {P.~T.}\ \bibnamefont
  {Cummings}},\ }\href {https://pubs.acs.org/doi/10.1021/jz201312e} {\bibfield
  {journal} {\bibinfo  {journal} {J. Phys. Chem. Lett.}\ }\textbf {\bibinfo
  {volume} {2}},\ \bibinfo {pages} {2859} (\bibinfo {year} {2011})}\BibitemShut
  {NoStop}%
\bibitem [{\citenamefont {Kondrat}\ \emph {et~al.}(2014)\citenamefont
  {Kondrat}, \citenamefont {Wu}, \citenamefont {Qiao},\ and\ \citenamefont
  {Kornyshev}}]{kondrat2014NM}%
  \BibitemOpen
  \bibfield  {author} {\bibinfo {author} {\bibfnamefont {S.}~\bibnamefont
  {Kondrat}}, \bibinfo {author} {\bibfnamefont {P.}~\bibnamefont {Wu}},
  \bibinfo {author} {\bibfnamefont {R.}~\bibnamefont {Qiao}},\ and\ \bibinfo
  {author} {\bibfnamefont {A.~A.}\ \bibnamefont {Kornyshev}},\ }\href
  {https://www.nature.com/articles/nmat3916} {\bibfield  {journal} {\bibinfo
  {journal} {Nat. Mater.}\ }\textbf {\bibinfo {volume} {13}},\ \bibinfo {pages}
  {387} (\bibinfo {year} {2014})}\BibitemShut {NoStop}%
\bibitem [{\citenamefont {P{\'e}an}\ \emph {et~al.}(2014)\citenamefont
  {P{\'e}an}, \citenamefont {Merlet}, \citenamefont {Rotenberg}, \citenamefont
  {Madden}, \citenamefont {Taberna}, \citenamefont {Daffos}, \citenamefont
  {Salanne},\ and\ \citenamefont {Simon}}]{pean2014dynamics}%
  \BibitemOpen
  \bibfield  {author} {\bibinfo {author} {\bibfnamefont {C.}~\bibnamefont
  {P{\'e}an}}, \bibinfo {author} {\bibfnamefont {C.}~\bibnamefont {Merlet}},
  \bibinfo {author} {\bibfnamefont {B.}~\bibnamefont {Rotenberg}}, \bibinfo
  {author} {\bibfnamefont {P.~A.}\ \bibnamefont {Madden}}, \bibinfo {author}
  {\bibfnamefont {P.-L.}\ \bibnamefont {Taberna}}, \bibinfo {author}
  {\bibfnamefont {B.}~\bibnamefont {Daffos}}, \bibinfo {author} {\bibfnamefont
  {M.}~\bibnamefont {Salanne}},\ and\ \bibinfo {author} {\bibfnamefont
  {P.}~\bibnamefont {Simon}},\ }\href
  {https://pubs.acs.org/doi/abs/10.1021/nn4058243} {\bibfield  {journal}
  {\bibinfo  {journal} {ACS Nano}\ }\textbf {\bibinfo {volume} {8}},\ \bibinfo
  {pages} {1576} (\bibinfo {year} {2014})}\BibitemShut {NoStop}%
\bibitem [{\citenamefont {Breitsprecher}\ \emph {et~al.}(2018)\citenamefont
  {Breitsprecher}, \citenamefont {Holm},\ and\ \citenamefont
  {Kondrat}}]{breitsprecher2018charge}%
  \BibitemOpen
  \bibfield  {author} {\bibinfo {author} {\bibfnamefont {K.}~\bibnamefont
  {Breitsprecher}}, \bibinfo {author} {\bibfnamefont {C.}~\bibnamefont {Holm}},\ and\ \bibinfo {author} {\bibfnamefont {S.}~\bibnamefont {Kondrat}},\ }\href
  {http://dx.doi.org/10.1021/acsnano.8b04785} {\bibfield  {journal} {\bibinfo
  {journal} {ACS Nano}\ }\textbf {\bibinfo {volume} {12}},\ \bibinfo {pages}
  {9733} (\bibinfo {year} {2018})}\BibitemShut {NoStop}%
\bibitem [{\citenamefont {Feng}\ \emph {et~al.}(2019)\citenamefont {Feng},
  \citenamefont {Chen}, \citenamefont {Bi}, \citenamefont {Goodwin},
  \citenamefont {Postnikov}, \citenamefont {Brilliantov}, \citenamefont
  {Urbakh},\ and\ \citenamefont {Kornyshev}}]{Guang2019PRX}%
  \BibitemOpen
  \bibfield  {author} {\bibinfo {author} {\bibfnamefont {G.}~\bibnamefont
  {Feng}}, \bibinfo {author} {\bibfnamefont {M.}~\bibnamefont {Chen}}, \bibinfo
  {author} {\bibfnamefont {S.}~\bibnamefont {Bi}}, \bibinfo {author}
  {\bibfnamefont {Z.~A.~H.}\ \bibnamefont {Goodwin}}, \bibinfo {author}
  {\bibfnamefont {E.~B.}\ \bibnamefont {Postnikov}}, \bibinfo {author}
  {\bibfnamefont {N.}~\bibnamefont {Brilliantov}}, \bibinfo {author}
  {\bibfnamefont {M.}~\bibnamefont {Urbakh}},\ and\ \bibinfo {author}
  {\bibfnamefont {A.~A.}\ \bibnamefont {Kornyshev}},\ }\href {\doibase 10.1103/PhysRevX.9.021024} {\bibfield  {journal} {\bibinfo  {journal} {Phys.
  Rev. X}\ }\textbf {\bibinfo {volume} {9}},\ \bibinfo {pages} {021024}
  (\bibinfo {year} {2019})}\BibitemShut {NoStop}%
\bibitem [{\citenamefont {Noh}\ and\ \citenamefont {Jung}(2019)}]{C8CP07200K}%
  \BibitemOpen
  \bibfield  {author} {\bibinfo {author} {\bibfnamefont {C.}~\bibnamefont
  {Noh}}\ and\ \bibinfo {author} {\bibfnamefont {Y.}~\bibnamefont {Jung}},\
  }\href {\doibase 10.1039/C8CP07200K} {\bibfield  {journal} {\bibinfo
  {journal} {Phys. Chem. Chem. Phys.}\ }\textbf {\bibinfo {volume} {21}},\
  \bibinfo {pages} {6790} (\bibinfo {year} {2019})}\BibitemShut {NoStop}%
\bibitem [{\citenamefont {Bi}\ \emph {et~al.}(2019)\citenamefont {Bi},
  \citenamefont {Chen}, \citenamefont {Wang}, \citenamefont {Feng},
  \citenamefont {Dinca}, \citenamefont {Kornyshev},\ and\ \citenamefont
  {Feng}}]{bi2019molecular}%
  \BibitemOpen
  \bibfield  {author} {\bibinfo {author} {\bibfnamefont {S.}~\bibnamefont
  {Bi}}, \bibinfo {author} {\bibfnamefont {M.}~\bibnamefont {Chen}}, \bibinfo
  {author} {\bibfnamefont {R.}~\bibnamefont {Wang}}, \bibinfo {author}
  {\bibfnamefont {J.}~\bibnamefont {Feng}}, \bibinfo {author} {\bibfnamefont
  {M.}~\bibnamefont {Dinca}}, \bibinfo {author} {\bibfnamefont {A.~A.}\
  \bibnamefont {Kornyshev}},\ and\ \bibinfo {author} {\bibfnamefont
  {G.}~\bibnamefont {Feng}},\ }\href
  {\doibase https://doi.org/10.1038/s41563-019-0598-7} {\bibfield  {journal} {\bibinfo
  {journal} {Nat. Mater.}\ }(\bibinfo {year} {2020})}\BibitemShut {NoStop}%
\bibitem [{\citenamefont {Chatterji}\ and\ \citenamefont
  {Horbach}(2007)}]{chatterji2007lb}%
  \BibitemOpen
  \bibfield  {author} {\bibinfo {author} {\bibfnamefont {A.}~\bibnamefont
  {Chatterji}}\ and\ \bibinfo {author} {\bibfnamefont {J.}~\bibnamefont
  {Horbach}},\ }\href {https://doi.org/10.1063/1.2431174} {\bibfield  {journal}
  {\bibinfo  {journal} {J. Chem. Phys.}\ }\textbf {\bibinfo {volume} {126}},\
  \bibinfo {pages} {064907} (\bibinfo {year} {2007})}\BibitemShut {NoStop}%
\bibitem [{\citenamefont {Asta}\ \emph {et~al.}(2019)\citenamefont {Asta},
  \citenamefont {Palaia}, \citenamefont {Trizac}, \citenamefont {Levesque},\
  and\ \citenamefont {Rotenberg}}]{Benjamin2019jcp}%
  \BibitemOpen
  \bibfield  {author} {\bibinfo {author} {\bibfnamefont {A.~J.}\ \bibnamefont
  {Asta}}, \bibinfo {author} {\bibfnamefont {I.}~\bibnamefont {Palaia}},
  \bibinfo {author} {\bibfnamefont {E.}~\bibnamefont {Trizac}}, \bibinfo
  {author} {\bibfnamefont {M.}~\bibnamefont {Levesque}},\ and\ \bibinfo
  {author} {\bibfnamefont {B.}~\bibnamefont {Rotenberg}},\ }\href {\doibase 10.1063/1.5119341} {\bibfield  {journal} {\bibinfo  {journal} {J. Chem.
  Phys.}\ }\textbf {\bibinfo {volume} {151}},\ \bibinfo {pages} {114104}
  (\bibinfo {year} {2019})}\BibitemShut {NoStop}%
\bibitem [{\citenamefont {Lian}\ \emph
  {et~al.}(2016{\natexlab{a}})\citenamefont {Lian}, \citenamefont {Zhao},
  \citenamefont {Liu},\ and\ \citenamefont {Wu}}]{Lian2016JCP}%
  \BibitemOpen
  \bibfield  {author} {\bibinfo {author} {\bibfnamefont {C.}~\bibnamefont
  {Lian}}, \bibinfo {author} {\bibfnamefont {S.}~\bibnamefont {Zhao}}, \bibinfo
  {author} {\bibfnamefont {H.}~\bibnamefont {Liu}},\ and\ \bibinfo {author}
  {\bibfnamefont {J.}~\bibnamefont {Wu}},\ }\href {\doibase 10.1063/1.4968037}
  {\bibfield  {journal} {\bibinfo  {journal} {J. Chem. Phys.}\ }\textbf
  {\bibinfo {volume} {145}},\ \bibinfo {pages} {204707} (\bibinfo {year}
  {2016}{\natexlab{a}})}\BibitemShut {NoStop}%
\bibitem [{\citenamefont {Babel}\ \emph {et~al.}(2018)\citenamefont {Babel},
  \citenamefont {Eikerling},\ and\ \citenamefont
  {L\"{o}wen}}]{babel2018impedance}%
  \BibitemOpen
  \bibfield  {author} {\bibinfo {author} {\bibfnamefont {S.}~\bibnamefont
  {Babel}}, \bibinfo {author} {\bibfnamefont {M.}~\bibnamefont {Eikerling}},\ and\ \bibinfo {author} {\bibfnamefont {H.}~\bibnamefont {L\"{o}wen}},\ }\href
  {https://pubs.acs.org/doi/10.1021/acs.jpcc.8b05559} {\bibfield  {journal}
  {\bibinfo  {journal} {J. Phy. Chem. C}\ }\textbf {\bibinfo {volume} {122}},\
  \bibinfo {pages} {21724} (\bibinfo {year} {2018})}\BibitemShut {NoStop}%
\bibitem [{\citenamefont {Jiang}\ \emph {et~al.}(2014)\citenamefont {Jiang},
  \citenamefont {Cao}, \citenamefont {Jiang},\ and\ \citenamefont
  {Wu}}]{jiang2014kinetic}%
  \BibitemOpen
  \bibfield  {author} {\bibinfo {author} {\bibfnamefont {J.}~\bibnamefont
  {Jiang}}, \bibinfo {author} {\bibfnamefont {D.}~\bibnamefont {Cao}}, \bibinfo
  {author} {\bibfnamefont {D.-e.}\ \bibnamefont {Jiang}},\ and\ \bibinfo
  {author} {\bibfnamefont {J.}~\bibnamefont {Wu}},\ }\href {\doibase 10.1021/jz5009533} {\bibfield  {journal} {\bibinfo  {journal} {J. Phys. Chem.
  Lett.}\ }\textbf {\bibinfo {volume} {5}},\ \bibinfo {pages} {2195} (\bibinfo
  {year} {2014})}\BibitemShut {NoStop}%
\bibitem [{\citenamefont {Brogioli}\ \emph {et~al.}(2011)\citenamefont
  {Brogioli}, \citenamefont {Zhao},\ and\ \citenamefont
  {Biesheuvel}}]{brogioli2011prototype}%
  \BibitemOpen
  \bibfield  {author} {\bibinfo {author} {\bibfnamefont {D.}~\bibnamefont
  {Brogioli}}, \bibinfo {author} {\bibfnamefont {R.}~\bibnamefont {Zhao}},\
  and\ \bibinfo {author} {\bibfnamefont {P.~M.}\ \bibnamefont {Biesheuvel}},\
  }\href {http://dx.doi.org/10.1039/C0EE00524J} {\bibfield  {journal} {\bibinfo
   {journal} {Energy Environ. Sci.}\ }\textbf {\bibinfo {volume} {4}},\
  \bibinfo {pages} {772} (\bibinfo {year} {2011})}\BibitemShut {NoStop}%
\bibitem [{\citenamefont {Lian}\ \emph
  {et~al.}(2016{\natexlab{b}})\citenamefont {Lian}, \citenamefont {Liu},
  \citenamefont {van Aken}, \citenamefont {Gogotsi}, \citenamefont
  {Wesolowski}, \citenamefont {Liu}, \citenamefont {Jiang},\ and\ \citenamefont
  {Wu}}]{lian2016enhancing}%
  \BibitemOpen
  \bibfield  {author} {\bibinfo {author} {\bibfnamefont {C.}~\bibnamefont
  {Lian}}, \bibinfo {author} {\bibfnamefont {K.}~\bibnamefont {Liu}}, \bibinfo
  {author} {\bibfnamefont {K.~L.}\ \bibnamefont {van Aken}}, \bibinfo {author}
  {\bibfnamefont {Y.}~\bibnamefont {Gogotsi}}, \bibinfo {author} {\bibfnamefont
  {D.~J.}\ \bibnamefont {Wesolowski}}, \bibinfo {author} {\bibfnamefont
  {H.~L.}~\bibnamefont {Liu}}, \bibinfo {author} {\bibfnamefont {D.~E.}~\bibnamefont
  {Jiang}},\ and\ \bibinfo {author} {\bibfnamefont {J.~Z.}~\bibnamefont {Wu}},\
  }\href {https://doi.org/10.1021/acsenergylett.6b00010} {\bibfield  {journal}
  {\bibinfo  {journal} {ACS Energy Lett.}\ }\textbf {\bibinfo {volume} {1}},\
  \bibinfo {pages} {21} (\bibinfo {year} {2016}{\natexlab{b}})}\BibitemShut
  {NoStop}%
\bibitem [{\citenamefont {Janssen}\ \emph {et~al.}(2017)\citenamefont
  {Janssen}, \citenamefont {Griffioen}, \citenamefont {Biesheuvel},
  \citenamefont {van Roij},\ and\ \citenamefont
  {Ern{\'e}}}]{janssen2017coulometry}%
  \BibitemOpen
  \bibfield  {author} {\bibinfo {author} {\bibfnamefont {M.}~\bibnamefont
  {Janssen}}, \bibinfo {author} {\bibfnamefont {E.}~\bibnamefont {Griffioen}},
  \bibinfo {author} {\bibfnamefont {P.~M.}~\bibnamefont {Biesheuvel}}, \bibinfo
  {author} {\bibfnamefont {R.}~\bibnamefont {van Roij}},\ and\ \bibinfo
  {author} {\bibfnamefont {B.~H.}~\bibnamefont {Ern{\'e}}},\ }\href {\doibase 10.1103/PhysRevLett.119.166002} {\bibfield  {journal} {\bibinfo  {journal}
  {Phys. Rev. Lett.}\ }\textbf {\bibinfo {volume} {119}},\ \bibinfo {pages}
  {166002} (\bibinfo {year} {2017})}\BibitemShut {NoStop}%
\bibitem [{\citenamefont {Ambrozevich}\ \emph {et~al.}(2018)\citenamefont
	{Ambrozevich}, \citenamefont {Ambrozevich}, \citenamefont {Sibatov},\ and\
	\citenamefont {Uchaikin}}]{Ambrozevich2018}%
\BibitemOpen
\bibfield  {author} {\bibinfo {author} {\bibfnamefont {A.~S.}\ \bibnamefont
		{Ambrozevich}}, \bibinfo {author} {\bibfnamefont {S.~A.}\ \bibnamefont
		{Ambrozevich}}, \bibinfo {author} {\bibfnamefont {R.~T.}\ \bibnamefont
		{Sibatov}}, \ and\ \bibinfo {author} {\bibfnamefont {V.~V.}\ \bibnamefont
		{Uchaikin}},\ }\href {\doibase 10.3103/S1068371218010029} {\bibfield
	{journal} {\bibinfo  {journal} {Russ. Electr. Eng.}\ }\textbf {\bibinfo
		{volume} {89}},\ \bibinfo {pages} {64} (\bibinfo {year} {2018})}\BibitemShut
{NoStop}%
\bibitem [{\citenamefont {Bazant}\ \emph {et~al.}(2004)\citenamefont {Bazant},
  \citenamefont {Thornton},\ and\ \citenamefont {Ajdari}}]{bazant2004diffuse}%
  \BibitemOpen
  \bibfield  {author} {\bibinfo {author} {\bibfnamefont {M.~Z.}\ \bibnamefont
  {Bazant}}, \bibinfo {author} {\bibfnamefont {K.}~\bibnamefont {Thornton}},\
  and\ \bibinfo {author} {\bibfnamefont {A.}~\bibnamefont {Ajdari}},\ }\href
  {http://dx.doi.org/10.1103/PhysRevE.70.021506} {\bibfield  {journal}
  {\bibinfo  {journal} {Phys. Rev. E}\ }\textbf {\bibinfo {volume} {70}},\
  \bibinfo {pages} {021506} (\bibinfo {year} {2004})}\BibitemShut {NoStop}%
\bibitem [{\citenamefont {Janssen}\ and\ \citenamefont
  {Bier}(2018)}]{janssen2018transient}%
  \BibitemOpen
  \bibfield  {author} {\bibinfo {author} {\bibfnamefont {M.}~\bibnamefont
  {Janssen}}\ and\ \bibinfo {author} {\bibfnamefont {M.}~\bibnamefont {Bier}},\
  }\href {\doibase 10.1103/PhysRevE.97.052616} {\bibfield  {journal} {\bibinfo
  {journal} {Phys. Rev. E}\ }\textbf {\bibinfo {volume} {97}},\ \bibinfo
  {pages} {052616} (\bibinfo {year} {2018})}\BibitemShut {NoStop}%
\bibitem [{\citenamefont {Palaia}(2019)}]{Palaia2019}%
\BibitemOpen
\bibfield  {author} {\bibinfo {author} {\bibfnamefont {I.}\
		\bibnamefont {Palaia}},} {PhD thesis},\
\bibinfo  {school} {Universit\'{e} Paris-Saclay} \bibinfo {year} {2019}{, (private communication)} \BibitemShut
{NoStop}%  
\bibitem [{\citenamefont {Janssen}(2019)}]{janssen2019II}%
  \BibitemOpen
  \bibfield  {author} {\bibinfo {author} {\bibfnamefont {M.}~\bibnamefont
  {Janssen}},\ }\href {\doibase 10.1103/PhysRevE.100.042602} {\bibfield
  {journal} {\bibinfo  {journal} {Phys. Rev. E}\ }\textbf {\bibinfo {volume}
  {100}},\ \bibinfo {pages} {042602} (\bibinfo {year} {2019})}\BibitemShut
  {NoStop}%
 \bibitem [{\citenamefont {Sakaguchi}\ and\ \citenamefont
  {Baba}(2007)}]{PhysRevE.76.011501}%
  \BibitemOpen
  \bibfield  {author} {\bibinfo {author} {\bibfnamefont {H.}~\bibnamefont
  {Sakaguchi}}\ and\ \bibinfo {author} {\bibfnamefont {R.}~\bibnamefont
  {Baba}},\ }\href {\doibase 10.1103/PhysRevE.76.011501} {\bibfield  {journal}
  {\bibinfo  {journal} {Phys. Rev. E}\ }\textbf {\bibinfo {volume} {76}},\
  \bibinfo {pages} {011501} (\bibinfo {year} {2007})}\BibitemShut {NoStop}% 
\bibitem [{\citenamefont {Biesheuvel}\ and\ \citenamefont
  {Bazant}(2010)}]{PhysRevE.81.031502}%
  \BibitemOpen
  \bibfield  {author} {\bibinfo {author} {\bibfnamefont {P.~M.}\ \bibnamefont
  {Biesheuvel}}\ and\ \bibinfo {author} {\bibfnamefont {M.~Z.}\ \bibnamefont
  {Bazant}},\ }\href {\doibase 10.1103/PhysRevE.81.031502} {\bibfield
  {journal} {\bibinfo  {journal} {Phys. Rev. E}\ }\textbf {\bibinfo {volume}
  {81}},\ \bibinfo {pages} {031502} (\bibinfo {year} {2010})}\BibitemShut
  {NoStop}%  
\bibitem [{\citenamefont {Zhao}\ \emph {et~al.}(2013)\citenamefont {Zhao},
  \citenamefont {Satpradit}, \citenamefont {Rijnaarts}, \citenamefont
  {Biesheuvel},\ and\ \citenamefont {van~der Wal}}]{zhao2013optimization}%
  \BibitemOpen
  \bibfield  {author} {\bibinfo {author} {\bibfnamefont {R.}~\bibnamefont
  {Zhao}}, \bibinfo {author} {\bibfnamefont {O.}~\bibnamefont {Satpradit}},
  \bibinfo {author} {\bibfnamefont {H.~H.~M.}~\bibnamefont {Rijnaarts}}, \bibinfo
  {author} {\bibfnamefont {P.~M.}\ \bibnamefont {Biesheuvel}},\ and\ \bibinfo
  {author} {\bibfnamefont {A.}~\bibnamefont {van~der Wal}},\ }\href
  {https://doi.org/10.1016/j.watres.2013.01.025} {\bibfield  {journal}
  {\bibinfo  {journal} {Water Res.}\ }\textbf {\bibinfo {volume} {47}},\
  \bibinfo {pages} {1941} (\bibinfo {year} {2013})}\BibitemShut {NoStop}%
\bibitem [{\citenamefont {Obliger}\ \emph {et~al.}(2014)\citenamefont
  {Obliger}, \citenamefont {Jardat}, \citenamefont {Coelho}, \citenamefont
  {Bekri},\ and\ \citenamefont {Rotenberg}}]{Benjamin2014pre}%
  \BibitemOpen
  \bibfield  {author} {\bibinfo {author} {\bibfnamefont {A.}~\bibnamefont
  {Obliger}}, \bibinfo {author} {\bibfnamefont {M.}~\bibnamefont {Jardat}},
  \bibinfo {author} {\bibfnamefont {D.}~\bibnamefont {Coelho}}, \bibinfo
  {author} {\bibfnamefont {S.}~\bibnamefont {Bekri}},\ and\ \bibinfo {author}
  {\bibfnamefont {B.}~\bibnamefont {Rotenberg}},\ }\href {\doibase 10.1103/PhysRevE.89.043013} {\bibfield  {journal} {\bibinfo  {journal} {Phys.
  Rev. E}\ }\textbf {\bibinfo {volume} {89}},\ \bibinfo {pages} {043013}
  (\bibinfo {year} {2014})}\BibitemShut {NoStop}%  
\bibitem [{\citenamefont {Pilon}\ \emph {et~al.}(2015)\citenamefont {Pilon},
  \citenamefont {Wang},\ and\ \citenamefont {d'Entremont}}]{pilon2015recent}%
  \BibitemOpen
  \bibfield  {author} {\bibinfo {author} {\bibfnamefont {L.}~\bibnamefont
  {Pilon}}, \bibinfo {author} {\bibfnamefont {H.}~\bibnamefont {Wang}},\ and\
  \bibinfo {author} {\bibfnamefont {A.~L.}\ \bibnamefont {d'Entremont}},\
  }\href {http://jes.ecsdl.org/content/162/5/A5158.full} {\bibfield  {journal}
  {\bibinfo  {journal} {J. Electrochem. Soc.}\ }\textbf {\bibinfo {volume}
  {162}},\ \bibinfo {pages} {A5158} (\bibinfo {year} {2015})}\BibitemShut
  {NoStop}%  
\bibitem [{\citenamefont {Lian}\ \emph {et~al.}(2019)\citenamefont {Lian},
  \citenamefont {Su}, \citenamefont {Li}, \citenamefont {Liu},\ and\
  \citenamefont {Wu}}]{cheng2019acsnano}%
  \BibitemOpen
  \bibfield  {author} {\bibinfo {author} {\bibfnamefont {C.}~\bibnamefont
  {Lian}}, \bibinfo {author} {\bibfnamefont {H.}~\bibnamefont {Su}}, \bibinfo
  {author} {\bibfnamefont {C.}~\bibnamefont {Li}}, \bibinfo {author}
  {\bibfnamefont {H.}~\bibnamefont {Liu}},\ and\ \bibinfo {author}
  {\bibfnamefont {J.}~\bibnamefont {Wu}},\ }\href {\doibase 10.1021/acsnano.9b03303} {\bibfield  {journal} {\bibinfo  {journal} {ACS
  Nano}\ }\textbf {\bibinfo {volume} {13}},\ \bibinfo {pages} {8185} (\bibinfo
  {year} {2019})}\BibitemShut {NoStop}%
 \bibitem [{\citenamefont {Eftekhari}(2019)}]{Ali2019}%
\BibitemOpen
\bibfield  {author} {\bibinfo {author} {\bibfnamefont {A.}~\bibnamefont
		{Eftekhari}},\ }\href {\doibase 10.1021/acssuschemeng.8b01075} {\bibfield
	{journal} {\bibinfo  {journal} {ACS Sustain. Chem. Eng.}\ }\textbf {\bibinfo
		{volume} {7}},\ \bibinfo {pages} {3692} (\bibinfo {year} {2019})}\BibitemShut
{NoStop}% 
\bibitem [{\citenamefont {H\o{}jgaard~Olesen}\ \emph
  {et~al.}(2010)\citenamefont {H\o{}jgaard~Olesen}, \citenamefont {Bazant},\
  and\ \citenamefont {Bruus}}]{PhysRevE.82.011501}%
  \BibitemOpen
  \bibfield  {author} {\bibinfo {author} {\bibfnamefont {L.}~\bibnamefont
  {H\o{}jgaard~Olesen}}, \bibinfo {author} {\bibfnamefont {M.~Z.}\ \bibnamefont
  {Bazant}},\ and\ \bibinfo {author} {\bibfnamefont {H.}~\bibnamefont
  {Bruus}},\ }\href {\doibase 10.1103/PhysRevE.82.011501} {\bibfield  {journal}
  {\bibinfo  {journal} {Phys. Rev. E}\ }\textbf {\bibinfo {volume} {82}},\
  \bibinfo {pages} {011501} (\bibinfo {year} {2010})}\BibitemShut {NoStop}%
  \bibitem [{\citenamefont {Boon}\ and\ \citenamefont {van
  Roij}(2011)}]{boon2011blue}%
  \BibitemOpen
  \bibfield  {author} {\bibinfo {author} {\bibfnamefont {N.}~\bibnamefont
  {Boon}}\ and\ \bibinfo {author} {\bibfnamefont {R.}~\bibnamefont {van
  Roij}},\ }\href {https://doi.org/10.1080/00268976.2011.554334} {\bibfield
  {journal} {\bibinfo  {journal} {Mol. Phys.}\ }\textbf {\bibinfo {volume}
  {109}},\ \bibinfo {pages} {1229} (\bibinfo {year} {2011})}\BibitemShut
  {NoStop}%
\bibitem [{\citenamefont {Allahyarov}\ \emph {et~al.}(2017)\citenamefont
  {Allahyarov}, \citenamefont {Löwen},\ and\ \citenamefont
  {Taylor}}]{ALLAHYAROV201773}%
  \BibitemOpen
  \bibfield  {author} {\bibinfo {author} {\bibfnamefont {E.}~\bibnamefont
  {Allahyarov}}, \bibinfo {author} {\bibfnamefont {H.}~\bibnamefont {Löwen}},\ and\ \bibinfo {author} {\bibfnamefont {P.~L.}\ \bibnamefont {Taylor}},\
  }\href {\doibase https://doi.org/10.1016/j.electacta.2017.04.158} {\bibfield
  {journal} {\bibinfo  {journal} {Electrochim. Acta}\ }\textbf {\bibinfo
  {volume} {242}},\ \bibinfo {pages} {73 } (\bibinfo {year}
  {2017})}\BibitemShut {NoStop}%
\bibitem [{\citenamefont {Janssen}\ and\ \citenamefont
  {Bier}(2019)}]{janssen2019transient}%
  \BibitemOpen
  \bibfield  {author} {\bibinfo {author} {\bibfnamefont {M.}~\bibnamefont
  {Janssen}}\ and\ \bibinfo {author} {\bibfnamefont {M.}~\bibnamefont {Bier}},\
  }\href {\doibase 10.1103/PhysRevE.99.042136} {\bibfield  {journal} {\bibinfo
  {journal} {Phys. Rev. E}\ }\textbf {\bibinfo {volume} {99}},\ \bibinfo
  {pages} {042136} (\bibinfo {year} {2019})}\BibitemShut {NoStop}%
\bibitem [{\citenamefont {Kouyzer}(2015)}]{kouyzer2015charging}%
 \BibitemOpen
  \bibfield  {author} {\bibinfo {author} {\bibfnamefont {P.~A.~J.}\
  \bibnamefont {Kouyzer}},\ } \href{https://dspace.library.uu.nl/handle/1874/316785} {Master's thesis},\
  \bibinfo  {school} {Utrecht University} \bibinfo {year} {2015}\BibitemShut
  {NoStop}%  
\bibitem [{\citenamefont {Gebbie}\ \emph {et~al.}(2017)\citenamefont {Gebbie},
	\citenamefont {Smith}, \citenamefont {Dobbs}, \citenamefont {Lee},
	\citenamefont {Warr}, \citenamefont {Banquy}, \citenamefont {Valtiner},
	\citenamefont {Rutland}, \citenamefont {Israelachvili}, \citenamefont
	{Perkin},\ and\ \citenamefont {Atkin}}]{susan2017}%
\BibitemOpen
\bibfield  {author} {\bibinfo {author} {\bibfnamefont {M.~A.}\ \bibnamefont
		{Gebbie}}, \bibinfo {author} {\bibfnamefont {A.~M.}\ \bibnamefont {Smith}},
	\bibinfo {author} {\bibfnamefont {H.~A.}\ \bibnamefont {Dobbs}}, \bibinfo
	{author} {\bibfnamefont {A.~A.}\ \bibnamefont {Lee}}, \bibinfo {author}
	{\bibfnamefont {G.~G.}\ \bibnamefont {Warr}}, \bibinfo {author}
	{\bibfnamefont {X.}~\bibnamefont {Banquy}}, \bibinfo {author} {\bibfnamefont
		{M.}~\bibnamefont {Valtiner}}, \bibinfo {author} {\bibfnamefont {M.~W.}\
		\bibnamefont {Rutland}}, \bibinfo {author} {\bibfnamefont {J.~N.}\
		\bibnamefont {Israelachvili}}, \bibinfo {author} {\bibfnamefont
		{S.}~\bibnamefont {Perkin}}, \ and\ \bibinfo {author} {\bibfnamefont
		{R.}~\bibnamefont {Atkin}},\ }\href {\doibase 10.1039/C6CC08820A} {\bibfield
	{journal} {\bibinfo  {journal} {Chem. Commun.}\ }\textbf {\bibinfo {volume}
		{53}},\ \bibinfo {pages} {1214} (\bibinfo {year} {2017})}\BibitemShut
{NoStop}%
\end{thebibliography}
\end{document}